\documentclass[preprints,article,accept,moreauthors,pdftex]{mdpi}

\usepackage{graphicx}
\usepackage{diagbox}

\firstpage{1}
\pubvolume{xx}
\issuenum{1}
\articlenumber{5}
\pubyear{2019}
\copyrightyear{2019}
\history{Received: date; Accepted: date; Published: date}

\pdfoutput=1

\Title{The XSTAR Atomic Database}

\Author{Claudio Mendoza $^{1,2}$\orcidA{},
Manuel A. Bautista $^{1}$\orcidB{},
J\'er\^ome Deprince $^{3}$\orcidC{},
Javier A. Garc\'ia $^{4}$\orcidD{}, \\
Efra\'in Gatuzz $^{5}$\orcidE{},
Thomas W. Gorczyca $^{1}$\orcidF{},
Timothy R. Kallman $^{6}$\orcidG{},
Patrick Palmeri $^{3}$\orcidH{}, \\
Pascal Quinet $^{3,7}$\orcidI{}
and Michael C. Witthoeft $^{6,8}$\orcidJ{}
}

\AuthorNames{Claudio Mendoza,
Manuel A. Bautista,
J\'er\^ome Deprince,
Javier A. Garc\'ia,
Efra\'in Gatuzz,
Thomas W. Gorczyca
Timothy R. Kallman,
Patrick Palmeri,
Pascal Quinet and
Michael C. Witthoeft
}

\address{%
$^{1}$ \quad Department of Physics, Western Michigan University, Kalamazoo, MI 49008, USA; claudio.mendozaguardia@wmich.edu; manuel.bautista@wmich.edu; thomas.gorczyca@wmich.edu \\
$^{2}$ \quad Venezuelan Institute for Scientific Research (IVIC), Caracas 1020, Venezuela\\
$^{3}$ \quad Physique Atomique et Astrophysique, Universit\'e de Mons -- UMONS, B-7000 Mons, Belgium; jerome.deprince@umons.ac.be; patrick.palmeri@umons.ac.be; pascal.quinet@umons.ac.be\\
$^{4}$ \quad Cahill Center for Astronomy and Astrophysics, California Institute of Technology, Pasadena, CA 91125, USA; javier@caltech.edu\\
$^{5}$ \quad Max Planck Institute for Extraterrestrial Physics, Gießenbachstraße 1, 85748, Garching bei M\"unchen, Germany\\
$^{6}$ \quad NASA Goddard Space Flight Center, Greenbelt, MD 20771, USA; timothy.r.kallman@nasa.gov; michael.c.witthoeft@nasa.gov\\
$^{7}$ \quad IPNAS, Universit\'e de Li\`ege, Sart Tilman, B-4000 Li\`ege, Belgium\\
$^{8}$ \quad ADNET Systems Inc., Bethesda, MD 20817, USA
}

\corres{Correspondence: claudio.mendozaguardia@wmich.edu; Tel.: +1-269-387-4935}

\abstract{We describe the atomic database of the \textsc{xstar} spectral modeling code, summarizing the systematic upgrades carried out in the past twenty years to enable the modeling of K lines from chemical elements with atomic number $Z\leq 30$ and recent extensions to handle high-density plasmas. Such plasma environments are found, for instance, in the inner region of accretion disks round compact objects (neutron stars and black holes), which emit rich information about the system physical properties. Our intention is to offer a reliable modeling tool to take advantage of the outstanding spectral capabilities of the new generation of X-ray space telescopes (e.g., {\sc xrism} and {\sc athena}) to be launched in the coming years. Data curatorial aspects are discussed and an updated list of reference sources is compiled to improve the database provenance metadata. Two \textsc{xstar} spin-offs---the \texttt{ISMabs} absorption model and the \texttt{uaDB} database---are also described.
}


\begin{document}

\section{Introduction}\label{intro}

The \href{https://heasarc.gsfc.nasa.gov/docs/software/xstar/xstar.html}{\sc xstar} code computes the physical conditions and emission spectra of a photoionized gas and has been widely used in astrophysics, most notably in X-ray astronomy, for the past 20 years \citep{kal01, bau01}. It assumes an ionizing radiation source surrounded by a spherical gas shell that absorbs and transfers parts of this radiation to finally emit a spectrum. From an input comprising the incident continuum, shell thickness, elemental abundances, and gas density, the code computes the ionization balance, level populations, and temperature to generate the gas opacity and the emitted line and continuum fluxes. The computational model determines simultaneously the gas state and radiation field at every point from the source. To ensure convergence in local thermodynamic equilibrium (LTE) conditions, it implements a detailed and consistent treatment of the radiative (bound--bound, bound--free, and free--free) and collisional (electron-impact excitation and ionization, electron--ion recombination, charge transfer and three-body recombination) processes, which in turn depends on the accuracy of the underlying database of atomic rate coefficients. Therefore, {\sc xstar} is an ideal platform to study data curation schemes for implementing and maintaining  application-based atomic databases.

\begin{figure}[H]
  \centering
  \includegraphics[width=0.7\textwidth]{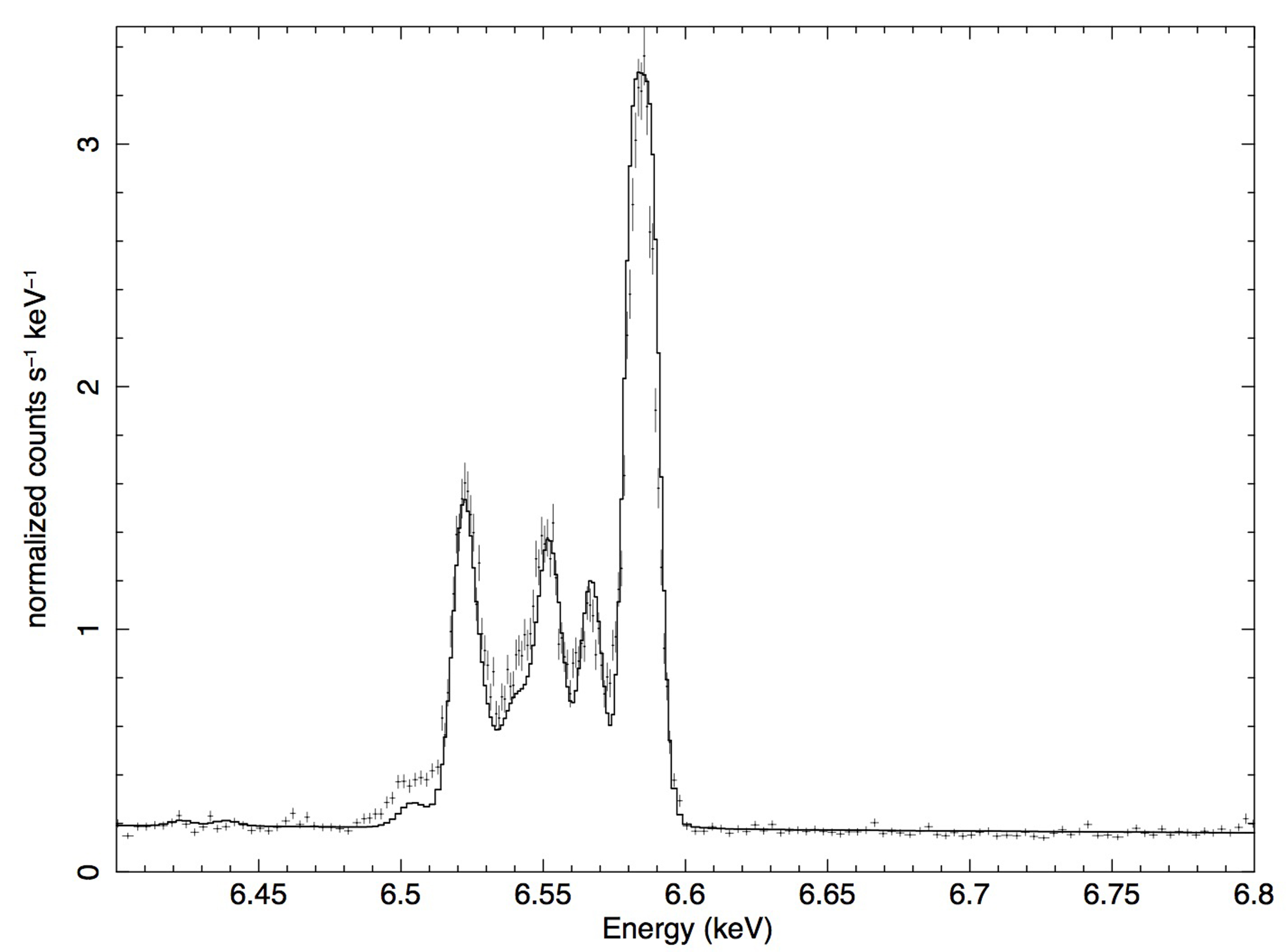}
  \caption{\textsc{hitomi} spectrum of the Perseus Cluster in the Fe~{\sc xxv} triplet spectral region (6.50--6.60~keV) \citep{hit16} showing the {\sc xstar} fit (black line). With the higher resolution, the iconic He-like triplet becomes a quartet and other Fe satellite lines begin to be detectable (e.g., the bump at $\sim 6.51$~keV). Reproduced from the XSTAR home page.}
  \label{perseus}
\end{figure}

Since 2001 we have been systematically computing for the {\sc xstar} database the atomic data required to model the K lines and edges in both cosmically abundant and trace elements \citep{pal02, pal03a, bau03, pal03b, pal03c, bau04, men04, gar05, jue06, pal08a, pal08b, wit09,  wit11a, wit11b, has10, pal11, gor13, has14, pal12, pal16, men17, men18, kal04, pal05, kal09, gar11, gat13a, gat13b, gat14, gat15, gat16, gat18a, gat19, gat20a, dep18, dep19a, dep19b, dep20, dep20b}. These X-ray spectral features are prime candidates to devise diagnostics to determine the plasma conditions and chemical abundances in a wide variety of astronomical entities: interstellar, intergalactic, and intra-cluster media; active galactic nuclei (AGN); X-ray binaries; supernova remnants; stars; the Sun, and some solar planets and satellites. This long-term collaboration has reached an important milestone, and the corresponding upgrading of the {\sc xstar} database with the new datasets has more than doubled its volume, bringing to the fore a series of data curational problems we discuss in this report. Our final goal is to groom {\sc xstar} to take advantage of the high spectral resolution (5~eV in the soft X-ray band $\sim 0.3{-}12$~keV) and sensitivity of the microcalorimeter-based spectrometers in the joint NASA/JAXA X-ray Imaging and Spectroscopy Mission (\href{https://heasarc.gsfc.nasa.gov/docs/xrism}{\sc xrism}) to be launched in 2022 and in the European Space Agency flagship Advanced Telescope for High-ENergy Astrophysics (\href{https://sci.esa.int/web/athena}{\sc athena}). As an example of the accuracy level we are committed to, Figure~\ref{perseus} shows the {\sc xstar} fit in the 6.50--6.60~keV region of the spectrum of the Perseus Cluster taken by the \href{https://www.nasa.gov/hitomi}{\sc hitomi} space telescope before its demise in 2016 \citep{hit16}.

We have recently introduced high-density ($n_e > 10^{18}$~cm$^{-3}$) effects in the {\sc xstar} database, which come into play in the reflection spectra of the inner region of the accretion disks round compact objects (e.g., black-hole candidates and neutron stars): continuum lowering \citep{dep18, dep19a, dep19b, dep20, dep20b}; dielectronic recombination (DR) suppression \citep{nik13, nik18}; collisional ionization; three-body recombination; and stimulated emission. Since the Fe abundance is a key measure of line reprocessing in reflection models \citep{gar16}, we are now in a good position to test if the neglect of such high-density effects is responsible for the anomalously high abundances recurrently derived  \citep{gar18}. Regarding {\sc xstar} database curation, the inclusion of high-density effects presents serious difficulties that are now managed by pre-processing the database at each density point. Since this pre-processing time is generally long, a grid of pre-computed databases has been deployed that, considering the data volumes involved, favors a cloud computing environment for the {\sc xstar} modeling of high-density plasmas.

Finally, the development of the {\sc xstar} database has given rise to two spin-offs: the \href{https://heasarc.gsfc.nasa.gov/xanadu/xspec/models/ismabs.html}{\texttt{ISMabs}} absorption model \citep{gat15} and the \href{https://heasarc.gsfc.nasa.gov/uadb/index.php}{\texttt{uaDB}} online database. The former is the result of benchmarks of the theoretical photoionization curves with observed K absorption in the interstellar medium (ISM), which has led to the implementation of a reliable model of ISM K absorption and to the matching of the theoretical K-line wavelengths with the astronomically observed and laboratory measured values, the latter two not always agreeing (e.g., O~{\sc i} \citep{kuh20}). The \texttt{uaDB} database, on  the other hand, allows the public downloading of the {\sc xstar} and other atomic rates without having to intervene the code.

\section{XSTAR Atomic Database}\label{database}

The {\sc xstar} database comprises records of multilevel model ions (usually referred to as ion targets) and data structures to derive the rates for the atomic microprocesses that determine the ionization and excitation of the plasma \citep{bau01}. The rate-type list is the following:

\begin{description}
\item[01.] Ground state ionization
\item[03.] Bound--bound collision
\item[04.] Bound--bound radiative
\item[05.] Bound--free collision (level)
\item[06.] Total recombination
\item[07.] Bound--free radiative (level)
\item[08.] Total recombination forcing normalization
\item[09.] Two-photon decay
\item[11.] Element data
\item[12.] Ion data
\item[13.] Level data
\item[14.] Bound--bound radiative, superlevel -- spectroscopic level
\item[15.] Collisional ionization total rate
\item[40.] Bound--bound collisional, superlevel -- spectroscopic level
\item[41.] Non-radiative Auger transition
\item[42.] Inner-shell photoabsorption followed by autoionization.
\end{description}

For each ionic species $(Z;N)$ in the ranges $1\leq Z\leq 30$ and $1\leq N\leq Z$, where $Z$ and $N$ are respectively the atomic and electron numbers, the data for the target models and processes are contained in ASCII files labeled $\mathtt{d}ion\mathtt{t}0dt$. The ion index $ion$ is defined as
\begin{equation}
    ion=Z(Z-1)/2+(Z -N)+1\ ,
\end{equation}
and $dt$ is the data type (see Appendix~\ref{A} for a complete list). Each $\mathtt{d}ion\mathtt{t}0dt$ file contains integer, floating-point, and character variables. For each release, the database is transcribed to a binary FITS-format \cite{gre03,pen10} derived structure consisting of three one-dimensional arrays (floating point, integer, and character) and a pointer array, which are uploaded in main memory at runtime. This scheme ensures portability, fast data uploading when the code is invoked, and single data accesses in main memory during processing.

The {\sc xstar} data curation strategy essentially follows two principles:
\begin{enumerate}
    \item Since the datasets are compiled from diverse sources, the original data structures and units are kept in the database records to be transcribed and unified at runtime. For instance, level energies are specified in eV but the photoionization cross sections are usually tabulated in Rydbergs. Levels in some atomic models are listed in intermediate coupling (fine structure) while in other $LS$ terms are adopted.  These format differences are accounted for by defining \textit{data types}, each of which adheres to its own rules for data tabulation.
    \item As new data and processes are included in the database, new data types may be defined. Thus, a photoionization cross sections can be specified in more than one data type---valence shell and K~shell, say---if the tabulation for each energy region comes from  different sources. They are then concatenated and adjusted at runtime.
\end{enumerate}

\begin{figure}[H]
  \centering
{\small
\begin{verbatim}
  6 13 0 4 6  8 0.0000000E+00  1.0000000E+00 ...  3 1 0 30  1 454 3s2.1S_0 %
  6 13 0 4 6 12 3.6651995E+01  1.0000000E+00 ...  3 3 1 30  2 454 3s1.3p1.3P_0 %
  6 13 0 4 6 12 3.8015000E+01  3.0000000E+00 ...  3 3 1 30  3 454 3s1.3p1.3P_1 %
  6 13 0 4 6 12 4.1638999E+01  5.0000000E+00 ...  3 3 1 30  4 454 3s1.3p1.3P_2 %
  6 13 0 4 6 12 5.6705005E+01  3.0000000E+00 ...  3 1 1 30  5 454 3s1.3p1.1P_1 %
  6 13 0 4 6 12 8.8177000E+01  1.0000000E+00 ...  3 3 1 30  6 454 3s0.3p2.3P_0 %
  6 13 0 4 6 12 8.9971999E+01  3.0000000E+00 ...  3 3 1 30  7 454 3s0.3p2.3P_1 %
  6 13 0 4 6 12 9.0909009E+01  5.0000000E+00 ...  3 3 1 30  8 454 3s0.3p2.3P_2 %
  6 13 0 4 6 12 9.5230001E+01  5.0000000E+00 ...  3 1 2 30  9 454 3s0.3p2.1D_2 %
  6 13 0 4 6 12 1.0956001E+02  1.0000000E+00 ...  3 1 0 30 10 454 3s0.3p2.1S_0 %
  6 13 0 4 6 16 1.0583000E+03  3.0000000E+00 ...  3 3 0 30 11 454 2p5.3s2.3p1.3S_1 %
  6 13 0 4 6 16 1.0607000E+03  5.0000000E+00 ...  3 3 2 30 12 454 2p5.3s2.3p1.3D_2 %
  6 13 0 4 6 16 1.0645000E+03  7.0000000E+00 ...  3 3 2 30 13 454 2p5.3s2.3p1.3D_3 %
  6 13 0 4 6 16 1.0654000E+03  3.0000000E+00 ...  3 1 1 30 14 454 2p5.3s2.3p1.1P_1 %
  6 13 0 4 6 16 1.0674000E+03  5.0000000E+00 ...  3 3 1 30 15 454 2p5.3s2.3p1.3P_2 %
  6 13 0 4 6 16 1.0766000E+03  1.0000000E+00 ...  3 3 1 30 16 454 2p5.3s2.3p1.3P_0 %
  6 13 0 4 6 16 1.0843000E+03  3.0000000E+00 ...  3 3 2 30 17 454 2p5.3s2.3p1.3D_1 %
  6 13 0 4 6 16 1.0895000E+03  3.0000000E+00 ...  3 3 1 30 18 454 2p5.3s2.3p1.3P_1 %
  6 13 0 4 6 16 1.0901000E+03  5.0000000E+00 ...  3 1 2 30 19 454 2p5.3s2.3p1.1D_2 %
  6 13 0 4 6 16 1.1029000E+03  1.0000000E+00 ...  3 1 0 30 20 454 2p5.3s2.3p1.1S_0 %
  6 13 0 4 6 20 1.2384461E+03  1.0000000E+00 ...  3 3 1 30 21 454 2s1.2p6.3s2.3p1.3P_0 %
  6 13 0 4 6 20 1.2391314E+03  3.0000000E+00 ...  3 3 1 30 22 454 2s1.2p6.3s2.3p1.3P_1 %
  6 13 0 4 6 20 1.2434606E+03  5.0000000E+00 ...  3 3 1 30 23 454 2s1.2p6.3s2.3p1.3P_2 %
  6 13 0 4 6 20 1.2457835E+03  3.0000000E+00 ...  3 1 1 30 24 454 2s1.2p6.3s2.3p1.1P_1 %
  6 13 0 4 6 24 9.7285003E+03  1.0000000E+00 ...  3 3 1 30 25 454 1s1.2s2.2p6.3s2.3p1.3P_0 %
  6 13 0 4 6 24 9.7293003E+03  3.0000000E+00 ...  3 3 1 30 26 454 1s1.2s2.2p6.3s2.3p1.3P_1 %
  6 13 0 4 6 24 9.7340004E+03  5.0000000E+00 ...  3 3 1 30 27 454 1s1.2s2.2p6.3s2.3p1.3P_2 %
  6 13 0 4 6 24 9.7373003E+03  3.0000000E+00 ...  3 1 1 30 28 454 1s1.2s2.2p6.3s2.3p1.1P_1 %
  6 13 0 4 6 10 6.6340128E+02  2.0000000E+00 ... 12 1 0 30 29 454 superlevel %
  6 13 0 4 6 14 1.0768748E+04  1.0000000E+00 ...  1 0 0 30 30 454 superlevel_[K] %
  6 13 0 4 6  9 6.9751000E+02  2.0000000E+00 ...  3 2 0 30 31 454 continuum %
\end{verbatim}
}
  \caption{Content of file $\mathtt{d}454\mathtt{t}006$ of the {\sc xstar} database listing the atomic model (data type $dt=06$ and rate type $rt=13$) for Mg-like Zn~{\sc xix} ($ion=454$).} \label{rec06}
\end{figure}

We show in Figure~\ref{rec06} the target model for Mg-like Zn~{\sc xix} complying with data type $dt=06$ and contained in file $\mathtt{d}454\mathtt{t}006$. Each row is headed by the integer tuple $(dt,rt,0,N_\mathrm{flt},N_\mathrm{int},N_\mathrm{chr})$, where $dt$ is the data type, $rt$ the rate type, $N_\mathrm{flt}$ the number of floating-point parameters, $N_\mathrm{int}$ number of integer parameters, and $N_\mathrm{chr}$ the length of the character string. As shown in the rate-type list above, $rt=13$ identifies level data. The first and second floating-point columns tabulate level energies (eV) and statistical weights $g=2J+1$ (the other two floating-point columns have been edited out). The second integer tuple $(n,2S+1,L,Z,i,ion)$ lists: the $n$ principal quantum number of the valence electron; the $2S+1$ spin multiplicity; the $L$ total orbital angular momentum; the $Z$ element atomic number; the $i$ level index, and the $ion$ identifier. The character string gives the level assignment (configuration and spectroscopic term), the end of the record being indicated with the $\%$ character. Records may span one arbitrary length line or multiple lines in the ASCII file.

This ion representation presents some remarkable features. By looking at the level $i$ index and assignment, the Zn~{\sc xix} model contains  data attributes for 31 levels from the valence shell ($1\leq i\leq 10$), open L shell ($11\leq i\leq 24$), and open K shell ($25\leq i\leq 28$), two superlevels ($i=29{-}30$), and a continuum level ($i=31$). The latter is associated to the ionization potential of the ion. Therefore, the model comprises spectroscopic bound levels, quasi-bound L- and K-vacancy resonances, and artificial levels. The tabulation must identify the ground level with index 1 and the continuum level with the highest index. Apart from this constraint, the levels are not necessarily listed in energy order facilitating model upgrading, which simply involves adding new levels at the end of the existing tabulation. The level-assignment string has been carefully devised and used throughout to facilitate searches based on filled shells, shell holes, $LS$ terms, and $J$ total angular momenta.

All other data types follow similar data structures (floating point, integer, and character) as those displayed in Figure~\ref{rec06}. For instance, we show in Figure~\ref{rec50} an excerpt of file $\mathtt{d}454\mathtt{t}050$ containing radiative data for the bound--bound transitions (upper level $k\leq 10$) in Zn~{\sc xix}. As before, the first integer tuple prescribes the data and rate types $dt=050$ and $rt=04$; three floating-point items; four integer items; and no character strings. The floating-point tuple $(\lambda, gf(i,k), A(k,i))$ lists the transition wavelength (\AA), oscillator strength ($gf$-value), and transition probability ($A$-value, s$^{-1}$). The second integer tuple  $(i,k,Z,ion)$ lists the lower and upper levels of the transition, atomic number, and ion identifier. By comparing the $i$ and $k$ transition indexes with the $i$ level indexes in Figure~\ref{rec06}, it may be seen that the tabulation includes both allowed and forbidden transitions; e.g., the resonance transition $(i,k)=(1,5)$ and the forbidden transition $(i,k)=(1,4)$. Importantly, every excited level in the database must include a decay mechanism, radiative and/or collisional, otherwise its population may grow out of bounds when modeling. The $\mathrm{3s1.3p1.3P\_0}$ metastable level ($i=2$) does not display a radiative transition ($j=0\rightarrow 0$ radiative transitions are strictly forbidden), and consequently, an alternative decay process must be provided in a separate data type; e.g., electron-impact collisional de-excitation.

\begin{figure}[H]
  \centering
{\small
\begin{verbatim}
           50 4 0 3 4 0   3.2047230E+02  5.2260000E-03  1.1310000E+08 1  3 30 454  %
           50 4 0 3 4 0   2.0629320E+02  7.0930000E-01  3.7060000E+10 1  5 30 454  %
           50 4 0 3 4 0   2.4417010E+02  2.5340000E-01  2.8350000E+10 3  6 30 454  %
           50 4 0 3 4 0   4.2222080E+02  4.7650000E-03  1.7830000E+08 5  6 30 454  %
           50 4 0 3 4 0   2.2582420E+02  2.7330000E-01  1.1910000E+10 2  7 30 454  %
           50 4 0 3 4 0   2.3171380E+02  1.9730000E-01  8.1690000E+09 3  7 30 454  %
           50 4 0 3 4 0   2.4845700E+02  3.1050000E-01  1.1180000E+10 4  7 30 454  %
           50 4 0 3 4 0   3.8631020E+02  1.4800000E-03  2.2060000E+07 5  7 30 454  %
           50 4 0 3 4 0   2.2265070E+02  3.4070000E-01  9.1680000E+09 3  8 30 454  %
           50 4 0 3 4 0   2.3806620E+02  7.8830000E-01  1.8560000E+10 4  8 30 454  %
           50 4 0 3 4 0   3.6175990E+02  1.2440000E-01  1.2680000E+09 5  8 30 454  %
           50 4 0 3 4 0   2.0084360E+02  2.0630000E-02  6.8230000E+08 3  9 30 454  %
           50 4 0 3 4 0   2.1330280E+02  2.0500000E-01  6.0100000E+09 4  9 30 454  %
           50 4 0 3 4 0   3.0751030E+02  8.4760000E-01  1.1960000E+10 5  9 30 454  %
           50 4 0 3 4 0   1.7136220E+02  2.6040000E-03  5.9150000E+08 3 10 30 454  %
           50 4 0 3 4 0   2.4339670E+02  3.8770000E-01  4.3650000E+10 5 10 30 454  %
           50 4 0 3 4 0   2.9315000E+02  0.0000000E+00  9.5940000E+00 1  4 30 454  %
\end{verbatim}
}
  \caption{Excerpt of file $\mathtt{d}454\mathtt{t}050$ of the {\sc xstar} database listing the radiative transitions with upper level $k\leq 10$ (data type $dt=50$ and rate type $rt=4$) for Mg-like Zn~{\sc xix} ($ion=454$).} \label{rec50}
\end{figure}

Data type $dt=86$, listing the radiative and Auger widths for K-vacancy levels, is also an interesting case as it involves transitions between different ion-charge states (see Figure~\ref{rec86}). The floating-point tuple $(E(k_N),A_a(k_N),A_a(k_N,i_{N-1}),A_r(k_N))$ lists: the  energy $E(k_N)$ (relative to the ionization threshold) and  total Auger width $A_a(k_N)$ of the $k_N$ level of the $N$-electron atom ($ion_N=454$); the partial Auger width $A_a(k_N,i_{N-1})$ leaving the ($N-1$)-electron ion ($ion_{N-1}=455$) in the $i_{N-1}$ level; and the total radiative width $A_r(k_N)$ of the $N$-electron ion. The second integer tuple then specifies $(i_{N-1},k_N, Z, ion_{N-1},ion_N)$. In this example the $(N-1)$-electron ion is assumed to end up in its ground state ($i_{N-1}=1$) after the Auger transition; however, this is not always the case as the Auger transition may involve an excited end level of the $(N-1)$-electron ion. Such partial Auger widths are not determined in some structure methods (e.g., {\sc hfr} \citep{cowan}) and in the $R$-matrix method \citep{ber95,burke} that includes radiative and Auger damping through an optical model potential \citep{gor99}.

\begin{figure}[H]
  \centering
{\small
\begin{verbatim}
 86 4 0 4 5 0   9.0309903E+03  1.1100000E+15  1.1100000E+15  1.1600000E+15 1 25 30 455 454 %
 86 4 0 4 5 0   9.0317903E+03  1.1000000E+15  1.1000000E+15  1.1800000E+15 1 26 30 455 454 %
 86 4 0 4 5 0   9.0364904E+03  1.1000000E+15  1.1000000E+15  1.1600000E+15 1 27 30 455 454 %
 86 4 0 4 5 0   9.0397903E+03  1.0800000E+15  1.0800000E+15  1.2900000E+15 1 28 30 455 454 %
\end{verbatim}
}
  \caption{Content of file $\mathtt{d}454\mathtt{t}086$ of the {\sc xstar} database listing the radiative and Auger widths (data type $dt=86$ and rate type $rt=4$) for Mg-like Zn~{\sc xix} ($ion=454$).} \label{rec86}
\end{figure}

Table~\ref{elev} in Appendix~\ref{B} tabulates the number of levels for each ionic species $(Z;N)$. It may be seen that most neutral and, in many cases, the singly ionized systems are not well represented, indicating that \textsc{xstar} is mainly destined to model photoionized plasmas for which such species have negligible fractions. The number of levels in the ion models is never above 1000 to enable practical runtimes when modeling, and the most relevant elements in the astronomical soft X-ray band---N, O, Ne, Fe, and Ne---are singularly well treated.

As fully described in the original paper on the \textsc{xstar} database \citep{bau01}, atomic datasets have been collected from a myriad of individual sources and data compilations such as the \href{https://www.nist.gov/pml/atomic-spectra-database}{NIST Atomic Spectra Database} \citep{kra20}, \href{http://cdsweb.u-strasbg.fr/topbase/topbase.html}{TOPbase} \citep{cun93}, \href{https://www.chiantidatabase.org/}{\sc chianti} \citep{der97}, \href{https://www.adas.ac.uk/}{ADAS}, \href{http://cdsweb.u-strasbg.fr/topbase/testop/TheIP.html}{Iron Project} \citep{hum93}, and \href{https://ned.ipac.caltech.edu/level5/Pradhan/Pradhan_contents.html}{PP95} \citep{pra95}. Major upgrades have been mainly concerned with the radiative and dielectronic recombination data from the tabulations in \href{http://amdpp.phys.strath.ac.uk/tamoc/DATA/}{AMDPP} \citep{bad03,bad06} and from the systematic computations of atomic data for K-line diagnostics summarized in Section~\ref{klines}. A complete data-provenance list is given in Appendix~\ref{C}.

\section{Computation of Atomic Data for K-line Diagnostics}\label{klines}

Since {\sc xstar} is widely used for spectral modeling in X-ray astrophysics, we have dedicated a great effort to compute the atomic data to characterize K lines in ionic species with $Z\leq 30$; namely, valence and K-vacancy level energies, transition wavelengths, $A$-values, radiative and Auger widths, and photoionization cross sections
\citep{bau03, pal03b, pal03c, bau04,men04, gar05, jue06, pal08a, pal08b,  wit09, wit11a, wit11b, has10, pal11, pal12, gor13, has14, pal16, men17, men18}. Due to the lack of spectroscopic data to validate most ionic models, a multi-code methodology was implemented right from the outset to compute the data and estimate their accuracy. Structure data were computed with the multiconfiguration codes {\sc hfr} (Pauli Hamiltonian) \citep{cow81}, {\sc autostructure} (Breit--Pauli Hamiltonian) \citep{eis74,bad11}, and {\sc grasp92} (Dirac--Coulomb--Breit Hamiltonian) \citep{par96} and the photoionization cross sections with the Breit--Pauli $R$-matrix suite of codes \citep{ber95,burke} and {\sc autostructure}. Due to large target models, the latter code was used to compute the photoionization cross sections in the distorted-wave approximation for species in isoelectronic sequences with electron number $19\leq N\leq 26$. Initial priority was given to ions from the oxygen, nitrogen, and iron isonuclear sequences and then from even-$Z$ elements, but those from odd-$Z$ and trace elements have been recently completed.

In the calculations of accurate atomic data for K-line diagnostics, two key effects must be considered in detail: radiative and Auger damping and orbital relaxation. In Sections~\ref{damping}--\ref{relaxation} we briefly go over these two processes and the methods to include them in structure and scattering calculations.

\begin{figure}[t]
  \centering
  \includegraphics[width=0.9\textwidth]{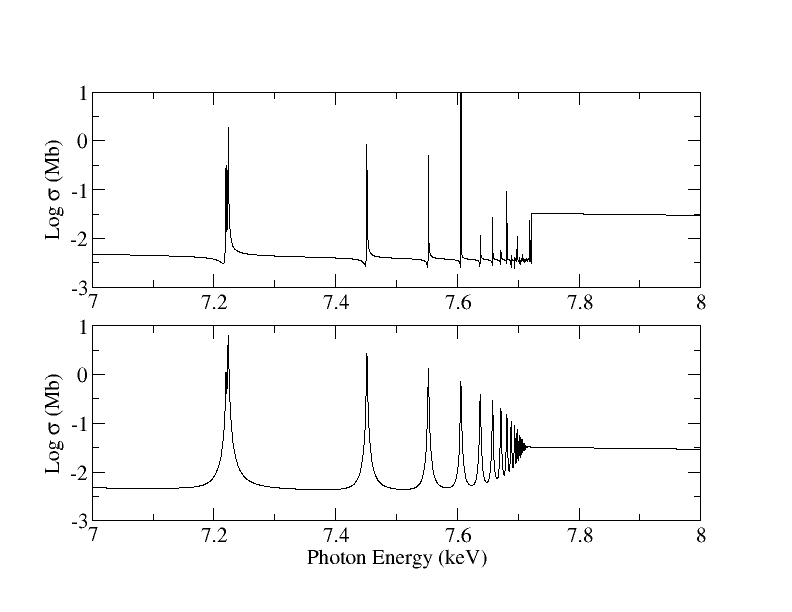}
  \caption{Photoabsorption cross section of the $\mathrm{1s^22s^22p^6\ ^1S}$ ground state of Ne-like Fe~{\sc xvii} in the K-edge region. {\it Top panel}: undamped cross section. {\it Bottom panel}: damped (radiation and Auger) cross section.}
  \label{fexvii}
\end{figure}

\subsection{Radiative and Auger Damping}\label{damping}

Let us consider an ion, say, with electron number $N\leq 10$. When a photon promotes a K-shell electron to an excited Rydberg state
\begin{equation}
h\nu+ \mathrm{1s}^2\mathrm{2s}^\lambda \mathrm{2p}^\mu\longrightarrow \mathrm{1s2s}^\lambda \mathrm{2p}^\mu n\mathrm{p}\ ,
\end{equation}
it subsequently decays via the radiative and Auger manifold
\begin{eqnarray}
\mathrm{1s2s}^\lambda \mathrm{2p}^\mu n\mathrm{p} & \stackrel{{\rm K}n}{\longrightarrow}
         \label{beta}        & \mathrm{1s}^2\mathrm{2s}^\lambda \mathrm{2p}^\mu + h\nu_n \\
         \label{alpha}       & \stackrel{{\rm K}\alpha}{\longrightarrow} &
                  \mathrm{1s}^2\mathrm{2s}^\lambda \mathrm{2p}^{\mu-1}n\mathrm{p}+h\nu_\alpha \\
         \label{part}     & \stackrel{{\rm KL}n}{\longrightarrow} &
                          \begin{cases}
                  \mathrm{1s}^2\mathrm{2s}^\lambda \mathrm{2p}^{\mu-1} + e^- \\
                  \mathrm{1s}^2\mathrm{2s}^{\lambda-1}\mathrm{2p}^\mu + e^-
                          \end{cases}  \\
         \label{spec}     & \stackrel{{\rm KLL}}{\longrightarrow} &
                          \begin{cases}
                  \mathrm{1s}^2\mathrm{2s}^\lambda \mathrm{2p}^{\mu-2} n\mathrm{p} + e^- \\
                  \mathrm{1s}^2\mathrm{2s}^{\lambda-1}\mathrm{2p}^{\mu-1}n\mathrm{p}   + e^- \\
                  \mathrm{1s}^2\mathrm{2p}^{\mu-2}n\mathrm{p} + e^-
                          \end{cases}
\end{eqnarray}
dominated by K$\alpha$ radiative  decay (Eq.~\ref{alpha}) and KLL Auger spectator-electron ionization (Eq.~\ref{spec}). Such channeling causes a damping effect (see Figure~\ref{fexvii}) since the $\mathrm{1s2s}^\lambda \mathrm{2p}^\mu n\mathrm{p}$ resonances have broad and symmetric widths almost independent of $n$, which must be taken into account when devising plasma diagnostics. In the $R$-matrix method it is computationally intractable to treat this channel array from first principles as it would involve target states with $n$p orbitals in an ever increasing close-coupling expansion. Therefore, for the higher Rydberg resonances, damping is neatly accounted for via an additional imaginary potential $V_\mathrm{opt}=-i(\Gamma_r+\Gamma_a)/2$ involving the total radiative width $\Gamma_r$---as developed in the Hickman--Robicheaux formalism \cite{hick,robicheaux95} (which is essentially equivalent to the Davies--Seaton formalism~\cite{davies})---and adapted to include the total Auger width $\Gamma_a$~\citep{gor99}.

\begin{figure}[t]
  \centering
  \includegraphics[width=0.9\textwidth]{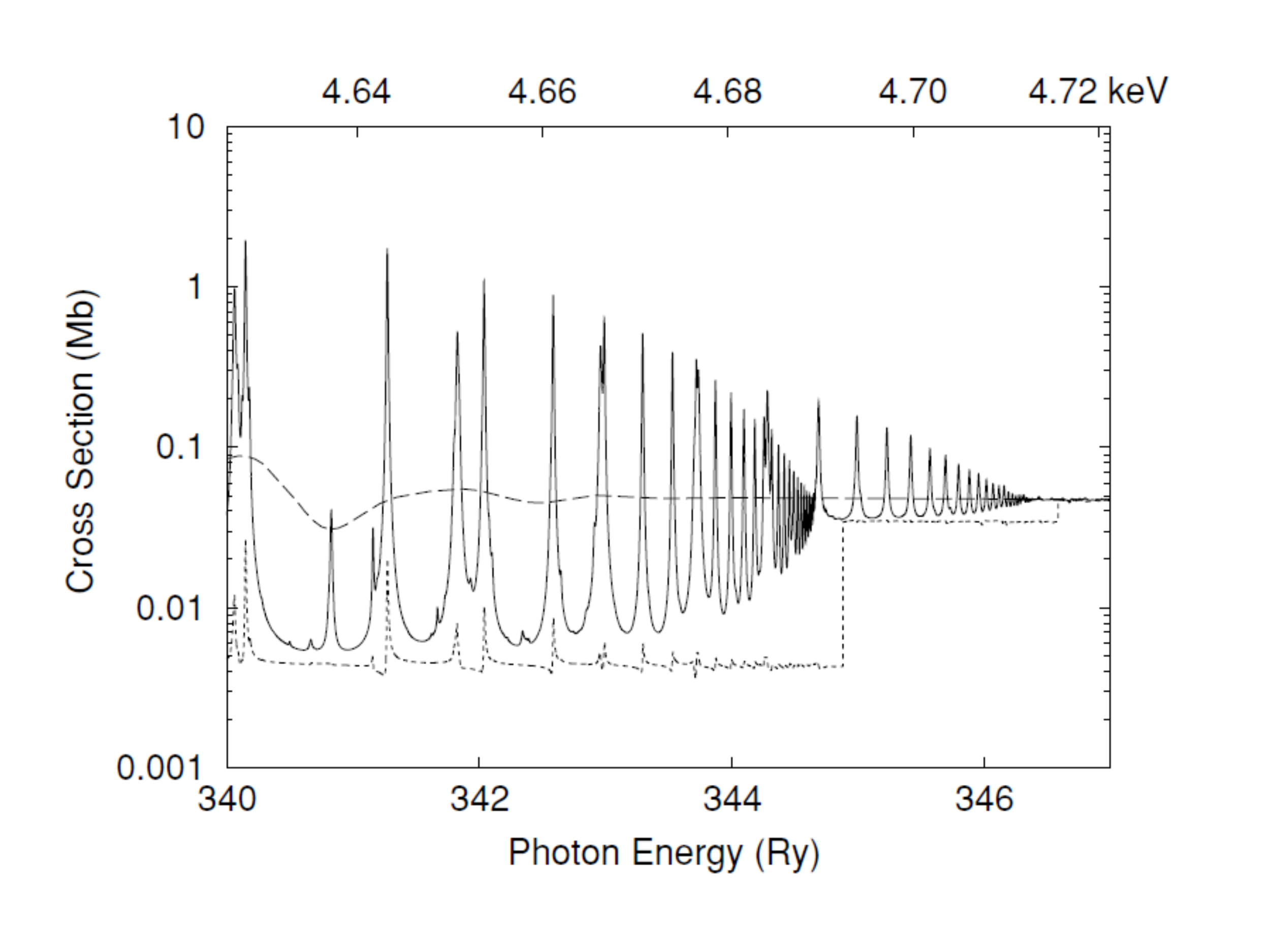}
  \caption{K-edge photoabsorption (full curve) and photoionization (dotted curve) cross sections of Ca~{\sc xv} (C-like) as a function of photon energy. The dashed curve gives the photoabsorption cross section convolved with a Gaussian of 0.001 width. Reproduced from Figure~3 of \citep{wit09} with permission of the AAS.}
  \label{clikeca}
\end{figure}

Due to their diagnostic importance, K$\alpha$ lines in the {\sc xstar} database are treated as bound--bound transitions and assigned wavelengths and $A$-values, but the upper K-vacancy states also display Auger widths that lead to autoionization (see Figures~\ref{rec06}--\ref{rec86}). However, the K resonances also appear in the photoabsorption cross sections as shown in Figure~\ref{fexvii} smearing the edge due to their peculiarly symmetric and constant line profiles. As a result, the K$\alpha$ resonances must be trimmed out from the partial photoionization cross sections that are used to derive the rates so as not to count them twice. However, it is more complicated to carry out this trimming procedure in the L-edge region due to the piling of a large number of resonances converging to closely spaced thresholds; thus some degree of double-counting is practically unavoidable making the L edge less useful to devise plasma diagnostics.

As shown in Figure~\ref{clikeca}, damping leads to a significant difference between photoabsorption and photoionization, and in the {\sc xstar} code both are needed: the former to determine opacities and the latter to derive the photoionization rates. However, the database essentially stores partial photoionization cross sections; i.e., cross sections leaving the photoionized $(N-1)$-electron system in specific states, and consequently, the difference between the total photoabsorption and photoionization cross sections is assigned to {\tt superlevel$\_$[K]} (see level $i=30$ in Figure~\ref{rec06}).

Due to the large number of target levels in ions with electron number $N>20$, their photoionization cross sections were computed using the simpler distorted-wave method implemented in {\sc autostructure} \citep{wit11b}. The level numbers for such ionic systems tabulated in Table~\ref{elev} (Appendix~\ref{B}) underwent considerable trimming to comply with the modeling specifications of \textsc{xstar}, and hence, do not represent the attributes of the actual collision targets. In {\sc autostructure},  photoabsorption is conveniently treated as the sum of two separate processes: photoionization and photoexcitation. However, the complicated decay routes of K-vacancy states delineated in Eqs.~(\ref{beta}--\ref{spec}) must be explicitly included in the photoexcitation process, and were thus computed in $LS$ coupling. Since \textsc{xstar} requires fine-structure partial photoionization cross sections to determine rates, we decided to obviate the phototexcitation process in ions with $Z\ne 26$ and $21\leq N\leq 26$ thus underestimating their opacities, although the radiative and Auger decays of the  L$\alpha,\beta$ and K$\alpha,\beta$ lines from these systems will be respectively taken into account in data types 50 and 86 (see Figures~\ref{rec50}--\ref{rec86}).

\subsection{Orbital Relaxation}\label{relaxation}

As shown in \citep{ffhf,cowan}, the photoionization of an inner-shell electron leads to a {\em relaxation} of the outer electrons of the ionized state due to a reduced screening of the atomic nuclear potential by the reduced inner electron cloud; i.e., an increased effective charge $Z_\mathrm{eff}$. For example, in the inner-shell photoionization of O~\textsc{i} $\mathrm{1s^22s^22p^4}$ to O~\textsc{ii} $\mathrm{1s2s^22p^4}$, the effective charge experienced by a 2p electron in the initial and final states is a complicated average screening including the other equivalent 2p electrons.  Nevertheless, by using hydrogenic considerations it is found that the electron orbital energy ($E_n\sim -E_R~Z_\mathrm{eff}^2/n^2$) is more tightly bound for the relaxed 2p electron, and the electron density is concentrated closer to the nucleus ($r_n\sim a_0~n^2/Z_\mathrm{eff}$).

\begin{figure}[t]
  \centering
  \includegraphics[width=0.8\textwidth]{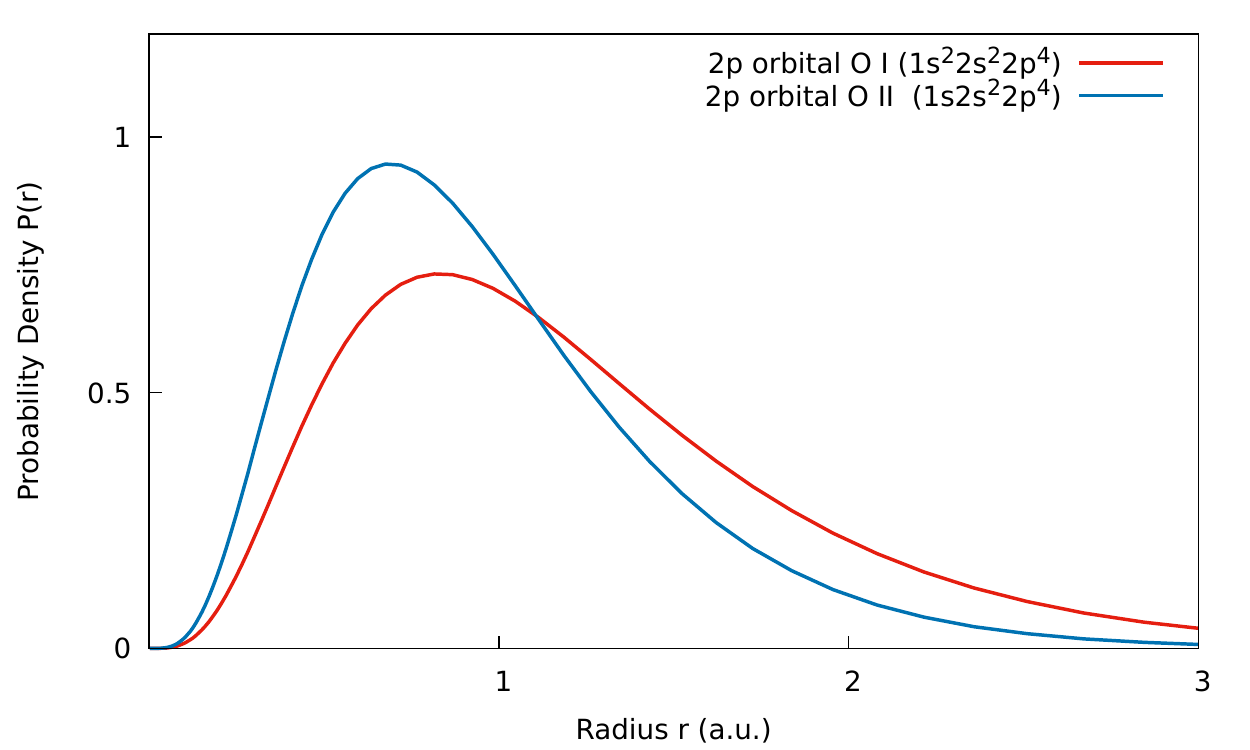}
  \caption{Probability density of the Hartree--Fock 2p orbital in the initial O~{\sc i} $\mathrm{1s^22s^22p^4}$ ground state compared to the 2p orbital in the final O~{\sc ii} $\mathrm{1s2s^22p^4}$ K-vacancy state. The latter 2p electrons are screened from the nucleus by one fewer inner-shell 1s electron, and are therefore ``relaxed'' to more tightly bound orbitals.}
  \label{figrelax}
\end{figure}

To illustrate this phenomenon, we show in Figure~\ref{figrelax} the outer 2p orbital in both neutral oxygen and its K-vacancy ionized daughter, obtained from single-configuration Hartree--Fock calculations \citep{ffmchf,mchf}.  Clearly the {\em relaxed} 2p orbital of the ionized O~{\sc ii} inner-shell state differs significantly from the neutral O~{\sc i} 2p orbital, a difference that needs to be accounted for in order to compute more accurate atomic transition properties.

In our {\sc hfr} and {\sc autostructure} structure calculations, orbital relaxation was taken into account by adopting different non-orthogonal orbital bases for each configuration. In most photoionization calculations, however, an orthonormal orbital basis is used because of the relative simplicity in computing transition and energy matrix elements. Relaxation then requires a substantial multi-configurational wave-function description including additional orbitals to account for the significant differences between initial- and final-state orbitals.  Thus, additional care must be taken in computing inner-shell photoionization cross sections when compared to outer-shell photoionization as discussed more fully in \citep{gor13}.


\section{Photoabsorption Cross-Section Benchmarks}

Table~\ref{prov} in Appendix~\ref{C} lists the reference sources for the computations of the photoabsorption cross sections. Measurement of K absorption in the laboratory and ISM enables useful benchmarks of the theoretical cross sections to improve spectral accuracy. In the following sections we give a summary of these procedures.

\subsection{Carbon Sequence}\label{Cions}

Calculations of the atomic data to model the K-shell photoabsorption by carbon ions were performed by \citep{has10}, and a benchmark was obtained using {\it Chandra} X-ray high-resolution spectra \citep{gat18a}. The C~{\sc ii} K$\alpha$ triplet as well as the C~{\sc iii} K$\alpha$ and K$\beta$ lines were identified in the X-ray spectra of four super-soft sources. However, C~{\sc i} was not modeled due to the presence of instrumental features at $\sim 43.6$~\AA.

\subsection{Oxygen Sequence}\label{Oions}

K-shell photoabsorption cross-sections for O~{\sc i} -- O~{\sc vii} were computed by \citep{gar05}, and comparisons of the theoretical resonance positions with astronomical observations and laboratory measurements were carried out by \citep{gat13a,gat13b}. The latter found that, due to a suspect experimental energy calibration scale, the X-ray astronomical observations provided a more reliable reference to perform the benchmark of the atomic data. For O~{\sc i}, the K-edge cross section was re-examined in detail by \citep{gor13} to yield a substantially improved version for the {\sc xstar} atomic database.

In an exhaustive study \citep{gor13}, the K$\alpha$ ($1s\rightarrow 2p$) resonance in O~{\sc i} was examined with $R$-matrix and \textsc{mchf} calculations, existing laboratory spectroscopic data, and several X-ray observational assessments as summarized in Table~\ref{olines}.  While practical $R$-matrix calculations for multi-electron systems are better tailored for computing scattering and atomic transition rates, they are less reliable in predicting the energy {\em differences} between atomic states as compared to limited bound-state structure calculations.  ($R$-matrix calculations necessarily include all quasi-bound and continuum final states, limiting a practical improvement the basis description.) The $R$-matrix K$\alpha$ energy \citep{gor13} was found to be $\sim 0.3$~eV higher than the final assessed observational value of 527.37~eV. More complex \textsc{mchf} calculations, which were focused only on the initial $\mathrm{1s^22s^22p^4}$ state and the final $\mathrm{1s2s^22p^5}$ state, converged downward to a value of 527.49~eV in good agreement with the observations.  On the other hand, the experiments performed at the Advanced Light Source (ALS) were reporting a value of 526.79~eV \citep{sto97}, a discrepancy of almost $-0.6$~eV.  Considering the reliability of the observational calibration and the convergence trend of the \textsc{mchf} calculations, the X-ray observation was unconventionally chosen instead of the laboratory measurement as the final resonance position in the atomic absorption model.

\begin{table}[H]
  \caption{Comparison of observed, measured, and computed O~{\sc i} K$\alpha$ and K$\beta$ line energies (eV).}\label{olines}
  \centering
  \begin{tabular}{lllll}
  \toprule
   \textbf{Method} & \textbf{Source} & $E(\mathrm{1s-2p})$ & $E(\mathrm{1s-3p})$ & $\Delta E$ \\
   \midrule
   Astronomical observations & {\it XMM-Neuwton}, Mkn~421 \citep{gor13}   & 527.30(5)  & 541.95(28)  & 14.65(33) \\
                             & {\it Chandra}, 7 sources \citep{gor13}     & 527.44(9)  & 541.72(18)  & 14.28(21) \\
                             & {\it Chandra}, shifted \citep{gor13}       & 527.26(9)  &             &           \\
                             & {\it Chandra}, 11 sources \citep{lia13}    & 527.39(2)  &             &           \\
                             & {\it Chandra}, 6 sources \citep{jue04}     & 527.41(18) & 541.77(40)  & 14.36(58) \\
   Laboratory measurements   & HZB \citep{leu20}                         & 527.26(4)  & 541.645(12) &           \\
                             & ALS \citep{mcl13a, mcl13b}                & 526.79(4)  & 541.19(4)   & 14.40(8)  \\
                             & ALS \citep{sto97}                         & 526.79(4)  & 541.20(4)   & 14.41(8)  \\
                             & WSRC \citep{men96}                        & 527.85(10) & 541.27(15)  & 13.41(25) \\
                             & Auger spectroscopy \citep{cal94}                              & 527.20(30) &             &           \\
    \textsc{mchf} converged result   & \textsc{mchf} \citep{gor13}                         & 527.49  &  &           \\
 \bottomrule
  \end{tabular}
\end{table}

A further consideration in assessing the reliability of the ALS measurements was that the absolute energy calibration relied on earlier molecular measurements with uncertainty estimates that seemed questionable.  Since the O~{\sc i} study by \citep{gor13}, new laboratory measurements have been performed using an electron beam ion trap (EBIT) \citep{leu20} to determine the K$\alpha$ resonance position at 527.6~eV, in good agreement with the observations and the converged \textsc{mchf} result. This new measurement suggests a recalibration of the ALS reference spectrum.

\subsection{Neon}\label{Neions}

For neon the {\sc xstar} atomic database includes the photoabsorption cross sections by \citep{jue06}. A benchmark for Ne~{\sc ii} and Ne~{\sc III} was performed with  {\it Chandra} and {\it XMM-Newton} observations of the bright low-mass X-ray binaries (LMXBs) Cygnus~X--2 and XTE~J1817-330 \citep{gat15}. Column densities were obtained for the cold ({\rm Ne}~{\sc i}), warm ({\rm Ne}~{\sc ii}, {\rm Ne}~{\sc iii}), and hot ({\rm Ne}~{\sc vii}, {\rm Ne}~{\sc ix}, {\rm Ne}~{\sc x}) components of the ISM.

\subsection{Magnesium}\label{Mgions}

New photabsorption cross sections for Mg species were computed by \citep{has14}. X-ray {\it XMM-Newton} spectra of the LMXB GS~1826-238 were used to perform a benchmark of the data. The resulting ionization fractions indicate that Mg is predominantly ionized rather than in neutral form as also shown by UV observations.

\subsection{Silicon}\label{Siions}

High-resolution {\it Chandra} spectra in the Si K-edge region (6$-$7 \AA) from 16 LMXBs were analyzed by \citep{gat20a}. The Si model included the {\rm Si}~{\sc i} photoabsorption cross sections computed by Gorczyca et al. (2020, in preparation), while those for the ionized species were taken from \citep{wit09}. The absorption features identified in the spectra agreed with the theoretical atomic data, even though the individual resonances of the K$\alpha$ triplet could not be resolved. Although a model without dust component was used, good data fits were obtained highlighting the need for accurate modeling of the gaseous component before attempting to address the solid component \citep{gat20a}.

\section{High-Density Effects on Atomic Parameters}

{\sc xstar} needs to address spectra from high-density sources such as the X-ray reflection from the inner region of the accretion disks round compact objects ($T_e\sim  10^5{-}10^7$~K and $n_e\sim 10^{18}{-}10^{22}$~cm$^{-3}$) \citep{gar18,fab10}. This requires taking into account a series of atomic processes usually neglected in plasma photoionization models, which have been fully discussed by \citep{kal20} and recently implemented in the code. We discuss in Sections~\ref{cl}--\ref{drs} two of these processes: continuum lowering and dielectronic recombination (DR) suppression, since they implied definite modifications of the database design and implementation.

\subsection{Continuum Lowering}\label{cl}

The density effects on the atomic thresholds (ionization potential and K edge), K$\alpha$ wavelengths and $A$-values, and Auger rates of oxygen and iron ions have been studied by \citep{dep18, dep19a, dep19b, dep20}. The atom is assumed to be embedded in a weakly coupled plasma represented in the \textsc{mcdf} method with a time-independent Debye--H\"uckel screened Dirac--Coulomb Hamiltonian
\begin{equation}
H^{DH}_{DC}=\sum_i c \vec{\alpha_i} \cdot \vec{p_i}+ \beta_i c^2 - \frac{Z}{r_i}
e^{-\mu r_i}
+ \sum_{i>j} \frac{1}{r_{ij}} e^{-\mu r_{ij}}\ ,
\label{dh}
\end{equation}
where $r_{ij}=|\vec{r}_i-\vec{r}_j|$ and the plasma screening parameter $\mu$ (inverse of the Debye shielding length $\lambda_D$) is given in atomic units by
\begin{equation}
\mu = \frac{1}{\lambda_D} = \sqrt{\frac{4\pi n_e}{k T_e}}\ .
\label{screen}
\end{equation}

\begin{figure}[t]
  \centering
  \includegraphics[width=0.53\textwidth]{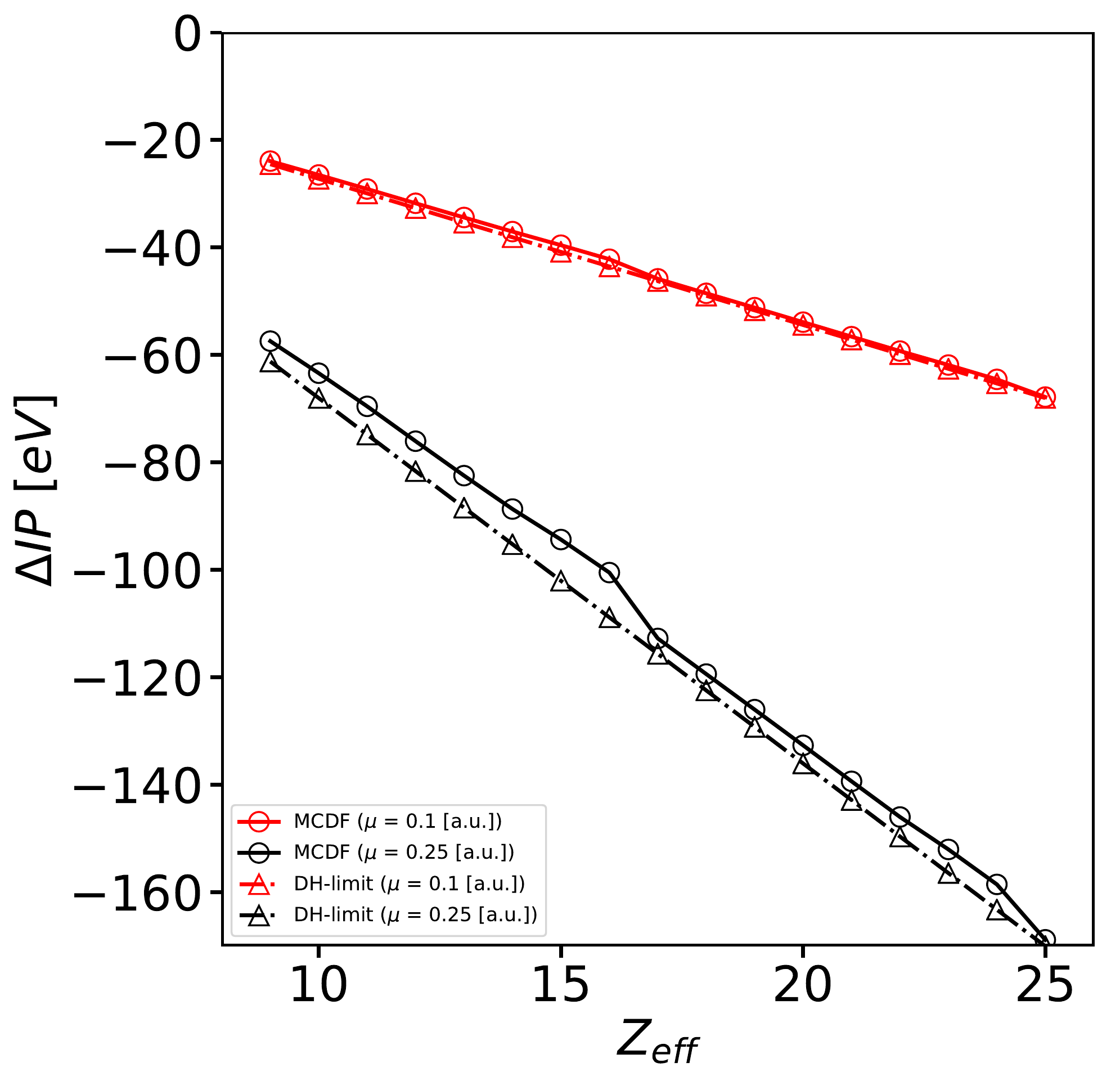}
  \caption{Ionization potential lowering in Fe~{\sc ix} -- Fe~{\sc xxv} as a function of effective charge $Z_\mathrm{eff}=Z-N+1$ obtained with the \textsc{mcdf} method using a Debye--H\"uckel screening potential. Red circles: $\mu=0.1$. Black circles: $\mu=0.25$. Triangles: Debye--H\"uckel limit $\Delta E_0^\mathrm{DH}=-27.2116\,\mu Z_\mathrm{eff}$. Reproduced from Figure~1 of \citep{dep20} with permission from Astronomy \& Astrophysics, $\copyright$ESO.}
  \label{conlow}
\end{figure}

The density effects mainly cause a threshold lowering that increases with $\mu$ (see Figure~\ref{conlow}), which for the ionization potential in eV can be approximated by the universal formula \citep{dep20c}
\begin{equation}
\Delta E_0=(-26.30\pm 0.08)\mu Z_\mathrm{eff}\ .
\end{equation}
In {\sc xstar}, continuum lowering leads to density-dependent targets with reduced numbers of levels, which must be derived at each density point causing a severe overhead at runtime. This bottleneck can be mitigated by pre-computing a $\mu$-grid of atomic databases but at the price of increasing the database volume significantly ($\sim 10$~GB), thus hampering the familiar package downloading procedure. Consequently, deployment strategies are being considered in terms of cloud virtual machines.

\subsection{DR Suppression}\label{drs}

\begin{figure}[H]
  \centering
  \includegraphics[width=1.0\textwidth]{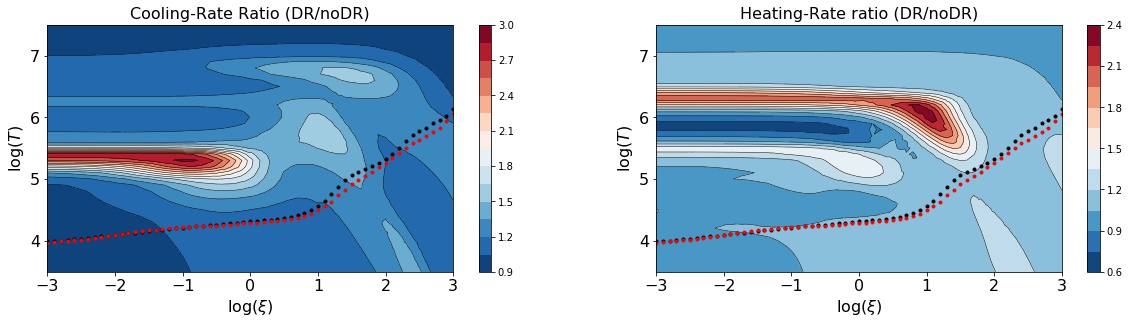}
  \caption{Cooling- and heating-rate ratios with and without DR obtained from an {\sc xstar} model with solar abundances, a power-law spectrum, density $n_e=10^{10}$~cm$^{-3}$, and luminosity $L=10^{32}$~erg\,s$^{-1}$. Black dotted curve: thermal temperature with DR.  Red dotted curve: thermal temperature without DR.}
  \label{heatcool}
\end{figure}

Rigorous collisional--radiative plasma modeling shows that DR is suppressed at high densities \citep{bad93}. To characterize DR plasma effects, we have compared the heating and cooling rates in the ionization-parameter--temperature plane
\begin{equation}
-3\leq \log(\xi)\leq 3~\mathrm{erg\,cm\,s}^{-1}\quad \mathrm{and}\quad 3.5\leq \log(T)\leq 7.5~\mathrm{K}
\end{equation}
for {\sc xstar} models with and without metal DR. We assume a power-law spectrum, density $n_e=10^{10}$~cm$^{-3}$, and luminosity $L=10^{32}$~erg\,s$^{-1}$ (see Figure~\ref{heatcool}). The larger temperature differences occur in the interval $1\leq\log(\xi)\leq 3$, where $T$ is lower by as much as 40\% when DR is excluded, and are caused by increases of the Fe average charge at $\log(\xi)\approx 1.3$ and $\log(\xi)\approx 2.5$. As shown in Figure~\ref{heatcool} ({\it right panel}), the DR/noDR heating-rate ratio reaches a maximum at $(\log(\xi),\log(T))\approx (1,6.3)$, where the dominant species with and without DR are Fe~{\sc xiii} and  Fe~{\sc xvii}, respectively. On the other hand, the cooling-rate ratio reaches a maximum at $(\log(\xi),\log(T))\approx (-1,5.4)$ caused by a higher oxygen average charge when DR is excluded.

\begin{figure}[H]
  \centering
  \includegraphics[width=0.49\textwidth]{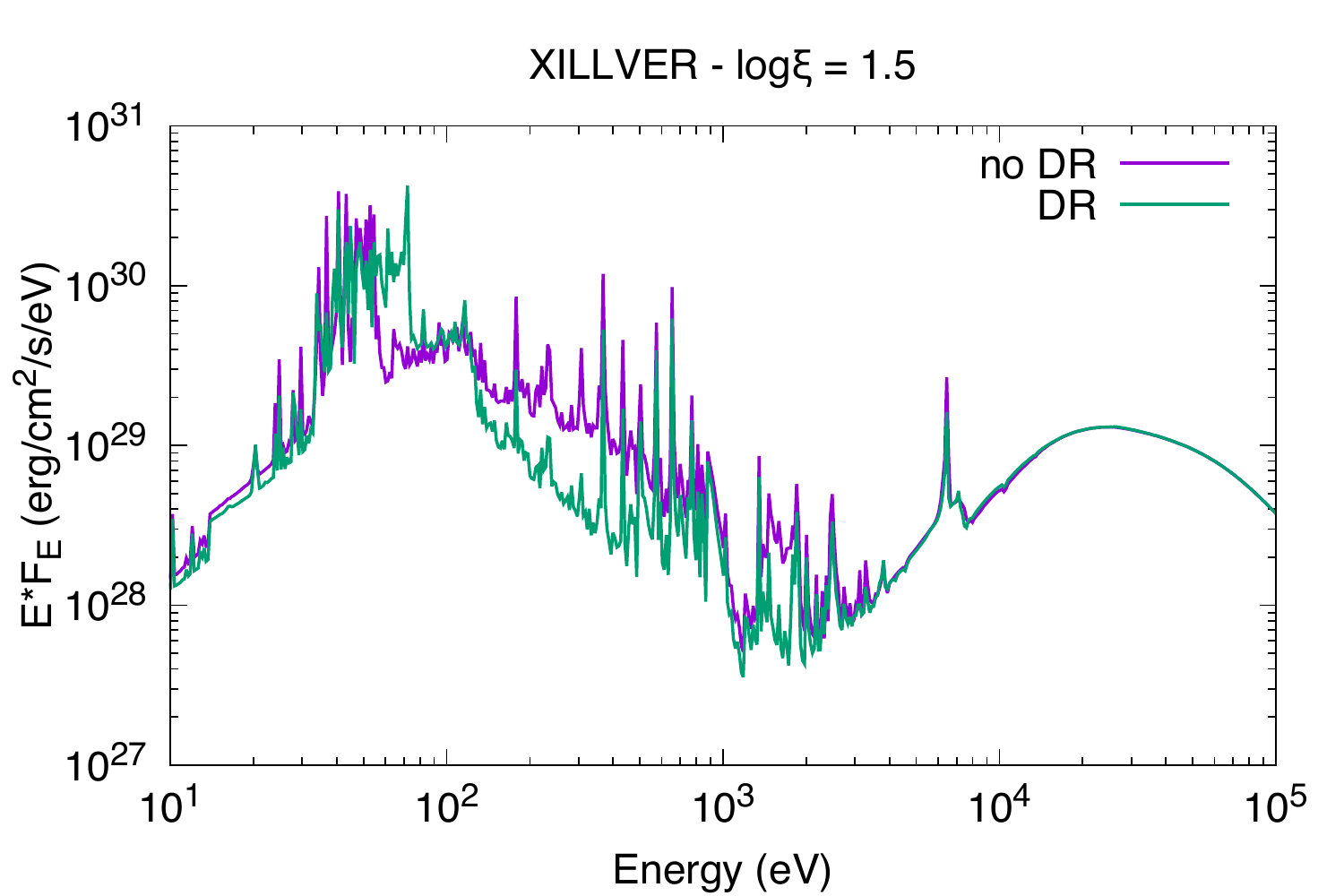}
  \includegraphics[width=0.49\textwidth]{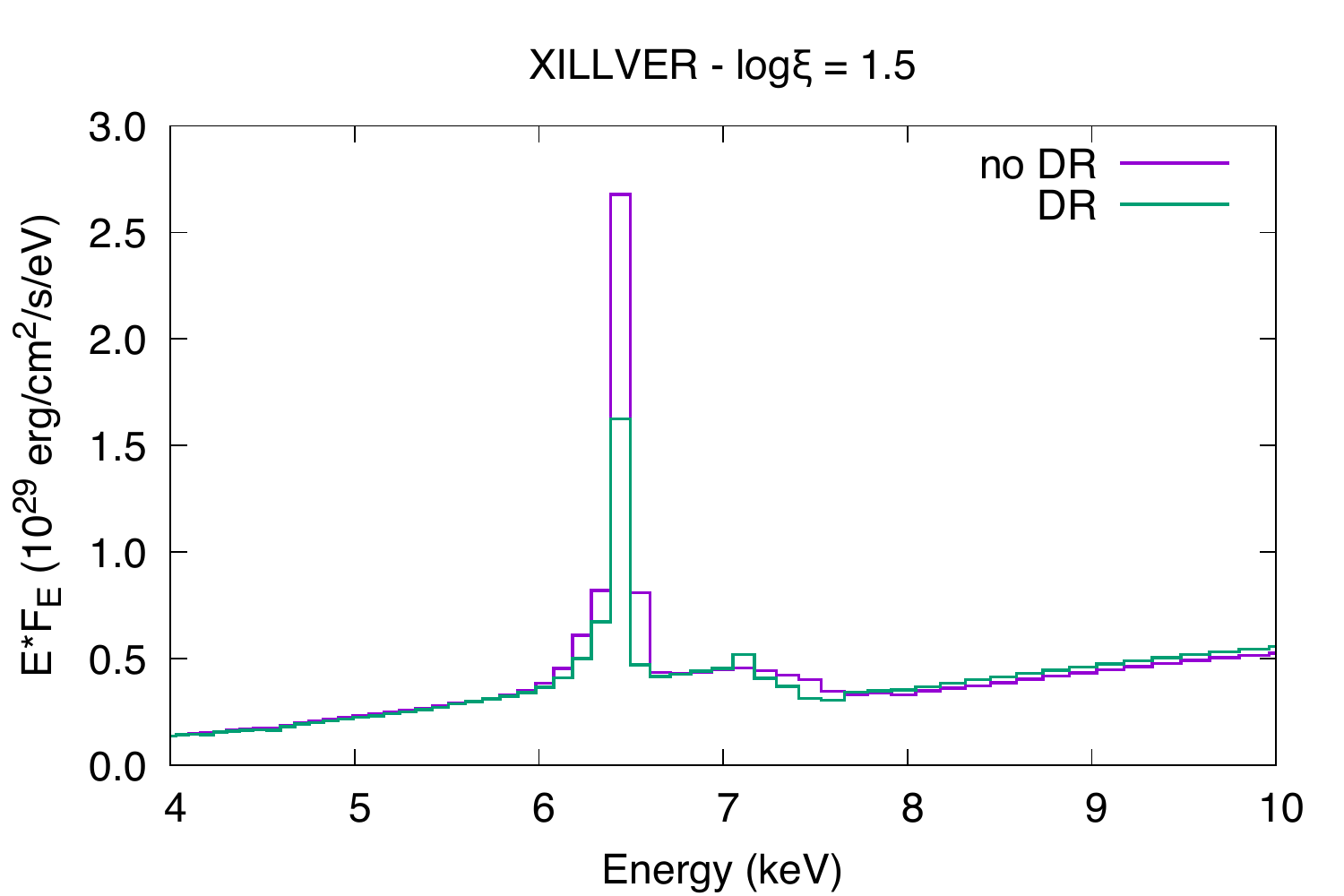}
  \caption{Reflected X-ray spectrum from an optically thick atmosphere modeled using {\sc xillver} with and without DR. A plane-parallel atmosphere with solar abundances and density $\log(n_e)=18$~cm$^{-3}$ was assumed.}
  \label{reflection}
\end{figure}

Concerning the Fe anomalous overabundance derived from the reflection spectra of accretion disks \citep{gar18}, we have computed the X-ray spectra reflected from an optically thick atmosphere using the code {\sc xillver} \citep{gar10, gar13}. The atmosphere is assumed to have plane-parallel geometry and a constant gas density of $\log(n_e)=18$~cm$^{-3}$. The ionization parameter is set to $\xi = 4\pi\,F_x/n_e = 32$~erg\,cm\,s$^{-1}$ implying an illuminating flux of $F_x = 3\times10^{18}$~erg\,cm$^{-2}$\,s$^{-1}$. The ionization balance is calculated self-consistently using the routines and atomic data from {\sc xstar} setting all elemental abundances to their solar values. It is found that metal DR suppression leads to significant changes in the ionization state of the gas resulting in more ionized species. As shown in Figure~\ref{reflection}, there is a strong enhancement of the Fe~{\sc xvii} L lines and  Fe~{\sc xviii} K lines.

It is worth mentioning that continuum lowering is not taken into account in this {\sc xillver} test calculation since it is expected to have a noticeable impact only at densities $n_e > 10^{19}$~cm$^{-3}$; the calculation was performed at $n_e > 10^{18}$~cm$^{-3}$ to show that DR effects become conspicuous at lower densities. Further characterization of high-density effects are reported in \citep{kal20}.

Density effects on DR and on the high-$n$ levels through continuum lowering are taken into account in the {\sc xstar} database via the rates into one or two fictitious superlevels for each ion. The recombination rate onto the superlevel is tabulated on a density--temperature grid and stored in data types 70 and 99 (see Appendix~\ref{A}).  For the former, rates are tabulated on a log scale while for the latter they are tabulated directly. The total recombination rate onto the ion is the sum of the rate onto the superlevels plus the rates onto the other levels that are treated explicitly; these are called the ``spectroscopic levels.'' The recombination rates onto the superlevels are chosen such that, when they are used to calculate a recombination rate and all the rates are summed, the total recombination rate for the superlevel(s) plus the spectroscopic levels adds to the total rate taken from one of the AMDPP compilations \citep{bad03} where available or from \citep{ald73} otherwise. We apply density-dependent suppression factors to the recombination  rates onto the superlevel to take into account density effects on radiative and dielectronic recombination.  They are derived from a general formula that depends on the isoelectronic sequence, charge, density, and temperature \citep{nik13, nik18}.

The superlevels  decay directly to the ground level without the emission of any observable cascade radiation for ions with three or more electrons. The exception is the decay of members of the H and He isoelectronic sequences for which we use explicitly calculated cascade matrix calculations to treat the decay of the superlevels to the spectroscopic levels \citep{bau98, bau00}. The superlevels are chosen to have energies close to the continuum. We include both radiative and collisional transitions to the ground level (and other levels in the case of H- and He-like ions).

\section{Database Curation}

The recent inclusion in the {\sc xstar} database of atomic parameters for odd-$Z$ elements and trace elements with $Z=(21{-}25,27,29,30)$ \citep{pal12, pal16, men17, men18} practically doubled the database volume when compared with previous versions that used hydrogenic scaling for these elements, bringing to the fore a series of curatorial problems. The database upgrade was thus lengthy and demanding in detail. In Sections \ref{atmodel}--\ref{metadata} we give a brief account of some the issues encountered and steps taken to solve them.

\subsection{Atomic Models}\label{atmodel}

As shown in Table~9 of \citep{pal12}, the target models for ions with atomic numbers $Z=(21{-}25,27,29,30)$ and electron number $19\leq N\leq Z-2$ include levels from the valence configurations $\mu=\mathrm{3d}^x\mathrm{4s}^y$  and the L- and K-vacancy configurations $\mathrm{3p}^{-1}\mu^{+1}$, $\mathrm{2p}^{-1}\mu^{+1}$, and $\mathrm{1s}^{-1}\mu^{+1}$. The total number of target levels in each of these species is larger than 500 and in some cases (e.g., $N=23$) close to 2000. It was found that to model astrophysical plasmas in {\sc xstar} with such  large atomic representations was impractical and with little tangible gains. We thus excluded the levels belonging to $\mu$ configurations containing 4s orbitals; specifically, $y=0$ in $\mu$ for $N>19$. This decision led to a reduction of the number of target levels to a few hundreds and implied extensive level renumbering for these sequences.

\subsection{Metastable Levels}

As discussed in Section~\ref{database}, the {\sc xstar} atomic models display occasional low-lying metastable levels (e.g., the $\mathrm{3s1.3p1.3P\_0}$ level in Mg-like ions) with no radiative decay transitions. For such levels, alternative decay mechanisms such as collisional de-excitation must be specified. However, for ions with electron number $N>19$ and ground configuration $\mathrm{3d}^x$, atomic models with a few hundred levels must be considered (see Section~\ref{atmodel}), and leads to the appearance of unexpected metastable levels at fairly high energies. Most of these levels were detected by running the code to look for levels with anomalously large populations, which were then removed by hand followed by level reordering.

\subsection{Database Pointers}\label{pointers}

As previously mentioned, the raw {\sc xstar} database consists of a list of ASCII files labeled $\mathtt{d}ion\mathtt{t}0dt$, where $ion$ is the ion identifier and $dt$ the data type. When the package is installed, these files are read into three one-dimensional arrays (integer, real, and character variables) and a set of pointers to be finally structured in a file under the binary FITS format. At runtime, these arrays are read into main memory to be accessed according to the user model. This data structure arrangement is efficient when uploading from disk and during processing, but can be cumbersome in database updating when new record types are created. Proficient nursing and testing are required before the new version is released.

\subsection{Metadata}\label{metadata}

The {\sc xstar} database compiles atomic data computed to order and from a myriad of other data sources. The data provenance of the first release was detailed in \citep{bau01}, but since then the database has undergone several mayor upgrades that have not been formally registered within its file structure.  One of the aims of the present paper is to create an up-to-date data provenance blueprint to be included in the database using the FITS metadata facilities.

In Table~\ref{prov} of Appendix~\ref{C} we have tabulated an up-to-date list of data source references including the original four- and five-letter mnemonics used for identification in \citep{bau01} and a new four-integer code to be inserted in every database record. For this purpose we adopt the third integer (currently displaying a zero) of the record heading tuple $(dt,rt,0,N_\mathrm{flt},N_\mathrm{int},N_\mathrm{chr})$ described in Section~\ref{database} to denote this provenance code.

\subsection{Relational Integrity}\label{integrity}

As described in Section~\ref{database}, the data in each $\mathtt{d}ion\mathtt{t}0dt$ file are self-contained to simplify and speedup the calculation of a wide variety of rates at runtime, but as a result this scheme obviates the relational integrity of the database. For instance, the wavelengths displayed in file $\mathtt{d}ion\mathtt{t}050$ (see Figure~\ref{rec50}) are not automatically recalculated if a level energy is updated in file $\mathtt{d}ion\mathtt{t}006$ (Figure~\ref{rec06}). The level attributes specified in the latter file are also repeated in several other data types; e.g., the K-vacancy level energy in file $\mathtt{d}ion\mathtt{t}086$ (Figure~\ref{rec86}). The extent of this complication is further illustrated in the list of data types in Appendix~\ref{A}. Data type 14 lists the ionization potential for every ionic species, but this datum is also contained in the floating-point tuple of every level in data type 06. Some of the level integer attributes of the latter data type are included in data types dealing with the effective ion charge (57), photoionization cross sections (49, 53, 85, and 88), superlevel recombination rates  (70 and 99), and satellite-level autoionization rates (72). This database model therefore can make updating an involved process that is certainly error prone.

\section{ISMabs}

Although the main components of the ISM are neutral ionic species, the also perceived existence of charged ions prompted us to develop a new ISM X-ray absorption model \citep{gat15} referred to as {\tt ISMabs}. It includes atomic data for singly and doubly ionized species for cosmically abundant elements, namely, H, He, C, N, O, Ne, Mg, Si, S, Ar, Ca, and Fe in addition to the neutral systems. This model allows ion column densities to be determined directly, and including ionized species has led to improved spectral fits when compared with those only considering neutral systems. The completeness and accuracy of the relevant atomic data are crucial to avoid misidentification and misinterpretation of the absorption features detected in analyses of ISM X-ray absorption spectra relying on fits based on Gaussian profiles. We have performed with this model a detailed study of the X-ray absorption in the local ISM by analyzing spectra from 24 Galactic sources. We have estimated the fractions of neutral and singly and doubly ionized species of O, Ne, and Fe confirming the dominance of the cold component and a pervasive low degree of ionization \citep{gat16}.

\section{Universal Atomic Database (uaDB)}

\begin{figure}[t]
\centering
\includegraphics[width=0.9\linewidth]{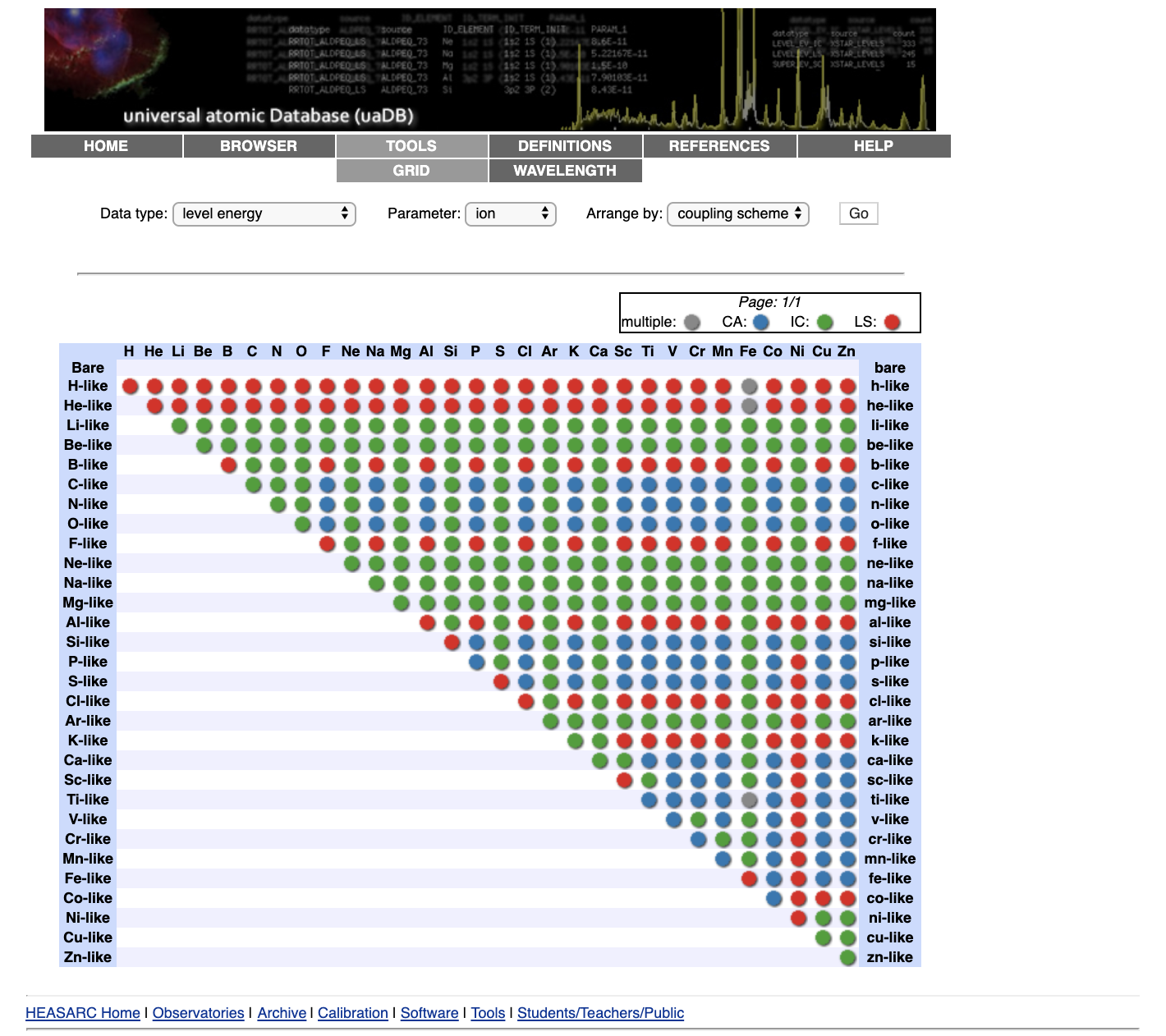}
\caption{Grid tool in \texttt{uaDB}.}
\label{fig:uadb_grid}
\end{figure}

The collection of atomic data in \textsc{xstar} is comprised of data in many different formats from many different sources. To assist in disseminating these data to the larger community and to account for the addition of new data, the Universal Atomic Database (\texttt{uaDB}) was developed. The \texttt{uaDB} is a mySQL database hosted at NASA/GSFC accessed via its \href{https://heasarc.gsfc.nasa.gov/uadb/index.php}{web site}, and data can be queried based on a number of properties such as ion and data type. The web page also contains tools to help assess what data are available across ions or data types. One such tool displays the type of data available on a grid of element \textit{vs.} ion (see Figure~\ref{fig:uadb_grid}) while another displays all atomic lines present in a given wavelength range.


The main features of \texttt{uaDB} are:

\begin{itemize}
\item Each dataset contains a reference to the source paper(s)
\item Data are stored in the original format and units as published
\item Multiple entries can exist for the same quantity (e.g., theoretical \textit{vs.}  measured energy levels)
\item The coupling scheme of atomic states is built into the database structure
\item Data collections exist that define a complete model; no duplicate data exist within a collection.
\end{itemize}

All data in \texttt{uaDB} have a reference to the data source. The reference consists of a link to the journal article where the data were obtained and to its entry in NASA/ADS. The user is never more than a couple mouse clicks away from reading the source article. Data are entered into \texttt{uaDB} in the same format and with the same units as published. This means that, for example, some energy levels are stored in eV and others in cm$^{-1}$. The \texttt{uaDB} understands these units and can internally convert them for data querying in an energy range. The \texttt{uaDB} can also convert between quantities that differ by a statistical weight (e.g., $f$-values and $gf$-values) since it understands the level structure. We do not modify published data in order to be as true as possible to the original source and to help identify transcription errors. However, this can make it more difficult to compare data from different sources. To help compare data, tools are developed that can perform the necessary conversions.

When compiling a set of atomic data for modeling purposes, it is important to have coverage; that is, for each state included in the model, a full set of data connecting that state with other states is necessary. If there are data allowing the population of a state to increase but no data to decrease it, then the model as a whole will be flawed. Furthermore, if the atomic model encompasses a large set of ions, it is difficult to obtain atomic data from the literature of equal quality. We often need to combine level-resolved data with term-resolved data to achieve coverage for an ion. To help with these difficulties, \texttt{uaDB} has different coupling schemes built into its structure and is aware of how different coupling schemes relate. Thus, when combining $LS$ states with $LSJ$, \texttt{uaDB} can detect if any states are being double-counted. There are also routines for converting data from one coupling scheme to another based on simple statistical weighting.

The \texttt{uaDB} currently contains the complete set of data included in \textsc{xstar} as well as the level data from \textsc{chianti}. To assist in distinguishing these data, \texttt{uaDB} defines collections. A collection is a dataset where there is no duplication of values for the same quantity. Currently, in \texttt{uaDB}, the collection is just a label that can be used to constrain a query, but in the future collections could be created dynamically to construct a custom model.

\section{Discussion and Conclusions}

The present report describes in detail the updated version of the underlying atomic database of the \textsc{xstar} photoionization code to replace the original implementation of \citep{bau01}. We include complete lists of the current rate types, data types (Appendix~\ref{A}), ionic models (Appendix~\ref{B}), and source references (Appendix~\ref{C}). We summarize the systematic calculations performed to characterize the metal K lines for astronomical X-ray spectral modeling and recent extensions of the database, namely, continuum lowering and dielectronic recombination suppression, to deal with high-density ($n_e\leq 10^{22}$~cm$^{-3}$) plasmas. We have made an attempt to convey the intricacies involved in the development and maintenance of an application-based atomic database, which go way beyond the compilation of the data leading to curatorial and integrity problems brought to the fore by the ever increasing dataset volumes.

An important finding of this work is that the atomic models used to computed the data are not necessarily suitable for plasma modeling, particularly for ionic species with electron number $N>20$. Level-number trimming was necessary to implement functional atomic representations for such systems although their soundness in non-LTE modeling deserves further attention as the appearance of unexpected metastable levels becomes an issue.

A further point worth emphasizing is that the  \textsc{xstar} atomic database and output files are structured under the FITS format specifications, which has been established as the standard for astronomical datasets particularly those associated to the new generation of telescopes. We believe that this compliance will reduce user stress in intense data processing, which is bound to take place at distributed data reservoirs rather than locally.

We have briefly recounted two projects that have emerged from the \textsc{xstar} database, namely the \texttt{ISMabs} absorption model and the \texttt{uaDB} database. The former has become an important initiative as it has led to benchmarks of the atomic data with observations and laboratory measurements; however, it has not been a smooth process as unexpected inaccuracies in the calibration of the laboratory wavelength scale were detected and finally resolved \citep{gor13, leu20}. This outcome has given us some confidence on the identification of the spectral features and the accuracy of our atomic data. The \texttt{uaDB} database model will become handy in future projects to disseminate larger atomic data volumes associated with non-LTE modeling.

A question that remains unanswered is whether our efforts to ensure accuracy and completeness for the \textsc{xstar} database are up to scratch to exploit the high resolution and sensitivity expected from the microcalorimeter-based spectrometers aboard the new X-ray space telescopes. The encouraging fit of the showcase \textsc{hitomi} spectrum depicted in Figure~\ref{perseus} required a certain amount of basic data manipulation, but at least it was manageable. Therefore, we envision further database refinement in the short future to provide a reliable modeling platform for high-resolution spectroscopy in a new era. In this respect more laboratory benchmarks would be welcome.

\vspace{6pt}



\authorcontributions{conceptualization, C.M., M.A.B., T.R.K.; methodology, C.M., E.G., J.A.G., J.D., M.A.B., M.C.W., P.P., P.Q., T.R.K., and T.W.G.; software, C.M., E.G., J.A.G., M.A.B., M.C.W., P.P., T.R.K., and T.W.G.; validation, J.A.G., M.A.B., P.P., T.R.K.; formal analysis, C.M., E.G., J.A.G., J.D., M.A.B., M.C.W., P.P., P.Q., T.R.K., and T.W.G.; investigation,  C.M., E.G., J.A.G., J.D., M.A.B., M.C.W., P.P., and, T.R.K.; validation, C.M., M.A.B., P.P., and T.R.K.; resources, M.A.B., P.Q., T.R.K.; data curation, C.M., M.A.B., P.P., and T.R.K; writing--original draft preparation, C.M., E.G., M.C.W., and T.R.K.; writing--review and editing, C.M., E.G., J.A.G., J.D., M.A.B., M.C.W., P.P., P.Q., T.R.K., and T.W.G.; visualization, C.M., M.C.W., and T.R.K; supervision, T.R.K.; project administration, M.A.B., P.Q., T.R.K.; funding acquisition, J.A.G., P.P., P.Q., and T.R.K.}

\funding{J.D. is a Research Fellow of the Belgian Fund for Research in Industry and Agriculture (FRIA). P.P. \& P.Q. are, respectively, Research Associate and Research Director of the Belgian Fund for Scientific Research F.R.S.--FNRS. Financial supports from these organizations as well as from the NASA Astrophysics Research and Analysis Program (grants 12-APRA12-0070; 80NSSC17K0345) are gratefully acknowledged.}

\acknowledgments{We are grateful to Dr. Anil. K. Pradhan (Ohio State University) for access to and space/time allocation at the Ohio Supercomputer Center and also to the NASA Private Cloud at GSFC for access and virtual machine installation, especially to Dr. Michael D. Moore for invaluable support.}

\conflictsofinterest{The authors declare no conflict of interest.}

\abbreviations{The following abbreviations are used in this manuscript:\\

\noindent
\begin{tabular}{@{}ll}
  ADAS  & Atomic Data and Analysis Structure \\
  ADS   & Astrophysics Data System \\
  ALS   &  Advanced Light Source \\
  AMDPP & Atomic and Molecular Diagnostic Processes in Plasmas \\
  FITS  & Flexible Image Transport System\\
  GSFC  & Goddard Space Flight Center \\
  HZB   & Helmholtz-Zentrum Berlin \\
  IPNAS & Institut de Physique Nucl\'eaire, Atomique et de Spectrom\'etrie\\
  LTE   & Local Thermodynamic Equilibrium \\
  MCHF  & Multi-Configuration Hartree--Fock \\
  NASA  & National Aeronautics and Space Administration\\
\end{tabular}
}

\noindent
\begin{tabular}{@{}ll}
  NIST  & National Institute of Standards and Technology \\
  PP95  & Atomic Data for the Analysis of Emission Lines \\
  WSRC  & Wisconsin Synchrotron Radiation Center
\end{tabular}

\appendix

\section{Data Types}\label{A}

\begin{description}
\item[01.] Radiative recombination rate coefficient of $N$-electron recombined ion \citep{ald73,ald76}: $\mathtt{r1} =A_\mathrm{rad}$~(cm$^3$\,s$^{-1}$); $\mathtt{r2}=\eta$; $\mathtt{i1}=ion_N$.
\item[02.] H$^0$ charge exchange rate coefficient of $N$-electron  recombined ion \citep{kin96}: $\mathtt{r1}= a$ ($10^{-9}$\,cm$^3$\,s$^{-1}$); $\mathtt{r2}= b$; $\mathtt{r3}= c$; $\mathtt{r4}= d$; $\mathtt{r5}= T_1$ (K); $\mathtt{r6}= T_2$ (K);  $\mathtt{r7}=\Delta E/k$ ($10^4$~K); $\mathtt{i1}=ion_N$; $\mathtt{s1}=$ recombining ion identifier.
\item[06.] Data attributes of the $i$th level of $N$-electron ion: $\mathtt{r1} =E(i)$~(eV); $\mathtt{r2} = (2J+1)$; $\mathtt{r3} = \nu$ (effective quantum number); $\mathtt{r4} =E(\infty)$~(eV); $\mathtt{i1} = n$; $\mathtt{i2}=(2S+1)$; $\mathtt{i3}=L$; $\mathtt{i4}=Z$; $\mathtt{i5}=i$; $\mathtt{i6}=ion_N$; $\mathtt{s1}=$ level configuration assignment.
\item[07.] Dielectronic recombination rate coefficient of $N$-electron recombined ion \citep{ald73,ald76}: $\mathtt{r1} =A_\mathrm{di}$~(cm$^3$\,s$^{-1}$\,K$^{3/2}$); $\mathtt{r2} =B_\mathrm{di}$; $\mathtt{r3}=T_0$~(K); $\mathtt{r4}=T_1$~(K); $\mathtt{i1}=ion_N$.
\item[14.] Ionization potential of $N$-electron ion:  $\mathtt{r1} =E(\infty)$~(eV);  $\mathtt{i1} = Z-N+1$;  $\mathtt{i2} = Z$; $\mathtt{i3}=ion_N$; $\mathtt{s1}=$ ion identifier.
\item[22.] Dielectronic recombination rate coefficient of the $N$-electron recombined ion \citep{nus86}: $\mathtt{r1}= a$; $\mathtt{r2}= b$; $\mathtt{r3}= c$; $\mathtt{r4}= d$; $\mathtt{r5}= e$; $\mathtt{r6}= f$; $\mathtt{i1}=ion_N$.
\item[30.] Total radiative recombination rate (hydrogenic) for $N$-electron recombined ion \citep{gou70}: $\mathtt{i1}=Z$; $\mathtt{i2}=ion_N$.
\item[38.] Total radiative recombination rate coefficient of $N$-electron recombined ion [http://amdpp.phys.strath.ac.uk/tamoc/DATA/RR/]: $\mathtt{r1} =A$ (cm$^3$\,s$^{-1})$; $\mathtt{r2} =B$; $\mathtt{r3} =T_0$ (K); $\mathtt{r4} =T_1$ (K); $\mathtt{r5} =C$; $\mathtt{r6} =T_2$ (K); $\mathtt{i1}=Z$; $\mathtt{i2}=N-1$; $\mathtt{i3}=M$; $\mathtt{i4}=W$; $\mathtt{i5}=ion_N$.
\item[39.] Total dielectronic recombination rate coefficient of $N$-electron recombined ion [http://amdpp.phys.strath.ac.uk/tamoc/DATA/DR/]: $\mathtt{r1{-}rj_{max}} =(C(j),j=1,j_\mathrm{max})$ (cm$^3$\,s$^{-1}$\,K$^{3/2}$); $\mathtt{rj_{max+1}{-}rj_{2*max}} =(T(j),j=1,j_\mathrm{max})$ (K); $\mathtt{i1}=Z$; $\mathtt{i2}=N-1$; $\mathtt{i3}=M$; $\mathtt{i4}=W$, $\mathtt{i5}=ion_N$.
\item[49.]  Partial photoionization cross section of $i_N$th level of the $N$-electron ion leaving the ($N-1$)-electron ion in the  $k_{N-1}$th level: $\mathtt{r1{-}rj_{2*max}} =(E(j),\sigma(E(j)),j=1,j_\mathrm{max})$~(Energy in Ryd relative to $E(\infty)$, cross section in Mb); $\mathtt{i1} = n$; $\mathtt{i2}=L$; $\mathtt{i3}=2J$; $\mathtt{i4}=Z$; $\mathtt{i5}=k_{N-1}$; $\mathtt{i6}=ion_{N-1}$; $\mathtt{i7}=i_N$; $\mathtt{i8}=ion_N$.
\item[50.] Line ($k-i$) radiation rates of $N$-electron ion: $\mathtt{r1} =\lambda$~(\AA); $\mathtt{r2} = gf(i,k)$; $\mathtt{r3} = A(k,i)$~(s$^{-1}$); $\mathtt{i1} = i$ (lower level); $\mathtt{i2}=k$ (upper level); $\mathtt{i3}=Z$; $\mathtt{i4}=ion_N$.
\item[51.]  Electron-impact effective collision strength for the $k - i$ transition of $N$-electron ion (CHIANTI fit \citep{bur92,der97}): $\mathtt{r1} =\Delta E$ (Ryd); $\mathtt{r2} =C$; $\mathtt{r3{-}r7} =(\Upsilon_{\rm red}(j),j=1,5)$ (reduced effective collision strength); $\mathtt{i1} = it$ (transition type); $\mathtt{i2} = i$ (lower level); $\mathtt{i3} = k$ (upper level); $\mathtt{i4} = Z$; $\mathtt{i5}=ion_N$.
\item[53.] TOPbase partial photoionization cross section (resonance averaged) of $i_N$th level of the $N$-electron ion leaving the ($N-1$)-electron ion in the  $k_{N-1}$th level: $\mathtt{r1{-}rj_{2*max}} =(E(j),\sigma(E(j)),j=1,j_\mathrm{max})$ (Energy in Ryd relative to $E(\infty)$, cross section in Mb); $\mathtt{i1} = n$; $\mathtt{i2}=L$; $\mathtt{i3}=2J$; $\mathtt{i4}=Z$; $\mathtt{i5}=k_{N-1}$; $\mathtt{i6}=ion_{N-1}$; $\mathtt{i7}=i_N$; $\mathtt{i8}=ion_N$.
\item[54.] Radiative transition probability $A_{ki}$ for the $k - i$ transition of $N$-electron ion computed by quantum defect theory (or hydrogenic): $\mathtt{r1} =0.0E+0$; $\mathtt{i1} = i$ (lower level); $\mathtt{i2}=k$ (upper level); $\mathtt{i3}=Z$; $\mathtt{i5}=ion_N$.
\item[56.] Electron-impact effective collision strengths for the $k - i$ transition of $N$-electron ion: $\mathtt{r1{-}rj_{max}} =(\log T_e(j), j=1, j_\mathrm{max})$~(K); $\mathtt{rj_{(max+1)}{-}rj_{(2*max)}} =(\Upsilon(T_e(j)),j=1,j_\mathrm{max})$ (effective collision strength); $\mathtt{i1} = i$ (lower level); $\mathtt{i2}=k$ (upper level); $\mathtt{i3}=Z$; $\mathtt{i5}=ion_N$.
\item[57.] Effective ion charge for $i$th level of $N$-electron ion: $\mathtt{r1} =Z_{\rm eff}$;  $\mathtt{i1} = n$; $\mathtt{i2}=L$; $\mathtt{i3}=2J$; $\mathtt{i4}=Z$; $\mathtt{i5}=i$;  $\mathtt{i6}=ion_N$
\item[59.] Partial photoionization cross section of $i_N$th level of the $N$-electron ion leaving the ($N-1$)-electron ion in the  $k_{N-1}$th level \citep{ver95}: $\mathtt{r1} =E(th)$~(eV); $\mathtt{r2} =E(0)$~(eV); $\mathtt{r3} =\sigma(0)$~(Mb); $\mathtt{r4} =y(a)$; $\mathtt{r5} =P$; $\mathtt{r6} =y(w)$; $\mathtt{i1} = N$; $\mathtt{i2}=n$ (shell principal quantum number); $\mathtt{i3}=l$ (orbital quantum number of the subshell); $\mathtt{i4}=k_{N-1}$; $\mathtt{i5}=ion_{N-1}$; $\mathtt{i6}=i_N$; $\mathtt{i7}=ion_N$; $\mathtt{s1}=$ shell--ion identifier.
\item[60.] Analytic fits for effective collision strengths in H-like ions \citep{cal94b}: $\mathtt{r1{-}rj_{max}} =$ coefficients; $\mathtt{i1} = i$ (lower level); $\mathtt{i2}=k$ (upper level); $\mathtt{i3}=1$; $\mathtt{i8}=ion_N$; $\mathtt{s1}=$ Transition.
\item[62.] Analytic fits for effective collision strengths in H-like ions \citep{cal94b}: $\mathtt{r1{-}rj_{max}} =$ coefficients; $\mathtt{i1} = i$ (lower level); $\mathtt{i2}=k$ (upper level); $\mathtt{i3}=1$; $\mathtt{i8}=ion_N$; $\mathtt{s1}=$ Transition.
\item[63.] Collisional transition probability $C_{ik}$ for $N$-electron ion computed by quantum defect theory (or hydrogenic): $\mathtt{i1} = 1$ $\mathtt{i2} = i$ (lower level); $\mathtt{i3}=k$ (upper level); $\mathtt{i4}=Z$; $\mathtt{i5}=ion_N$.
\item[66.] Fits to fine-structure collision strengths for He-like ions \citep{kat89}: $\mathtt{r1{-}rj_{max}} =$ coefficients; $\mathtt{i1} = i$ (lower level); $\mathtt{i2}=k$ (upper level); $\mathtt{i3}=Z$; $\mathtt{i4}=ion_N$.
\item[67.]  Analytic fits for effective collision strengths in He-like ions \citep{kee87b}: $\mathtt{r1{-}rj_{max}} =$ coefficients; $\mathtt{i1} = i$ (lower level); $\mathtt{i2}=k$ (upper level); $\mathtt{i3}=Z$; $\mathtt{i4}=ion_N$.
\item[68.]  Analytic fits for effective collision strengths in He-like ions \citep{zha87}: $\mathtt{r1{-}rj_{max}} =$ coefficients; $\mathtt{i1} = i$ (lower level); $\mathtt{i2}=k$ (upper level); $\mathtt{i3}=Z$; $\mathtt{i8}=ion_N$.
\item[69.] Fits to $LS$ collision strengths for He-like ions \citep{kat89}: $\mathtt{r1{-}rj_{max}} =$ coefficients; $\mathtt{i1} = i$ (lower level); $\mathtt{i2}=k$ (upper level); $\mathtt{i3}=Z$; $\mathtt{i8}=ion_N$.
\item[70.]  Coefficients for recombination and photoionization cross sections of superlevels: $\mathtt{r1{-}rj_{nd}} =(n_e(j),j=1,j_{nd})$; $\mathtt{rj_{nd+1}{-}rj_{nd+nt}} =(T_e(j),j=1,j_{nt})$; $\mathtt{rj_{nd+nt+1}{-}rj_{ nd+nt+nt*nd}} =((\log\alpha(j,j'),j'=1,j'_{nd}),j=1,j_{nt})$; $\mathtt{rj_{nd+nt+nt*nd+1}-rj_{nd+nt+nt*nd+2*nx}} =(E(j),\sigma(j),j=1,j_{nx})$; $\mathtt{i1} = nd$; $\mathtt{i2} = nt$; $\mathtt{i3} = nx$; $\mathtt{i4} = n$; $\mathtt{i5}=L$; $\mathtt{i6}=2S+1$; $\mathtt{i7}=Z$; $\mathtt{i8}=k_{N-1}$; $\mathtt{i9}=ion_{N-1}$; $\mathtt{i10}=i_N$; $\mathtt{i11}=ion_N$.
\item[71.] Radiative transition rates from superlevels to spectroscopic levels: $\mathtt{r1{-}rj_{nd}} =(n_e(j),j=1,j_{nd})$; $\mathtt{rj_{nd+1}{-}rj_{nd+nt}} =(T_e(j),j=1,j_{nt})$; $\mathtt{rj_{nd+nt+1}{-}rj_{ nd+nt+nt*nd}} =((A(j,j'),j'=1,j'_{nd}),j=1,j_{nt})$; $\mathtt{r_{nd+nt+nt*nd+1}} =\lambda$~(\AA); $\mathtt{i1} = nd$; $\mathtt{i2} = nt$;  $\mathtt{i3} = i$ (lower level); $\mathtt{i4} = k$ (upper level); $\mathtt{i5}=Z$;  $\mathtt{i6}=ion_N$.
\item[72.] Autoionization rates for satellite levels: $\mathtt{r1} = A_a(k,i)$~(s$^{-1}$);  $\mathtt{r2} =E(k)$~(eV above ionization limit); $\mathtt{r3} = (2J+1)$;  $\mathtt{i1}=(2S+1)$; $\mathtt{i2}=L$; $\mathtt{i3}=k$ (level); $\mathtt{i4}=i$ (continuum level); $\mathtt{i5}=Z$;  $\mathtt{i6}=ion_N$; $\mathtt{s1}=$ level configuration.
\item[73.] Fit to effective collision strengths for satellite levels of He-like ions \citep{sam83}: $\mathtt{r1{-}rj_{7}} = $ fit coefficients; $\mathtt{i1}=i$ (lower level);  $\mathtt{i2}=j$ (upper level); $\mathtt{i3}=Z$;  $\mathtt{i4}=ion_N$.
\item[74.] Delta functions to add to photoionization cross sections to match ADF DR rates: $\mathtt{r1} = E(\infty)$~(eV); $\mathtt{r1{-}rj_{m}} =(E(j),j=1,j_{m})$ (eV); $\mathtt{rj_{m+1}{-}rj_{2m}} =(f(j),j=1,j_{m})$ (cm$^2$); $\mathtt{i1} = n$; $\mathtt{i2}=L$; $\mathtt{i3}=2S+1$; $\mathtt{i4}=Z$; $\mathtt{i5}=k_{N-1}$; $\mathtt{i6}=ion_{N-1}$; $\mathtt{i7}=i_N$; $\mathtt{i8}=ion_N$
\item[75.] Autoionization rates for  Fe~{\sc xxiv} satellites \citep{bau03}: $\mathtt{r1} = A_a(k,i)$~(s$^{-1}$);  $\mathtt{r2} =E(k)$~(eV above ionization limit);  $\mathtt{i1}=ion_N$, $\mathtt{i2}=k_{N}$; $\mathtt{i3}=ion_{N-1}$; $\mathtt{i4}=i_{N-1}$; $\mathtt{i5}=ion_N$.
\item[76.] Two-photon radiation rate for ($k-i$) transition of $N$-electron ion: $\mathtt{r1} = A(k,i)$~(s$^{-1}$); $\mathtt{i1} = i$ (lower level); $\mathtt{i2}=k$ (upper level); $\mathtt{i3}=1$; $\mathtt{i4}=ion_N$; $\mathtt{s1}=$ transition identifier.
\item[77.] Collision transition rates from superlevels to spectroscopic levels: $\mathtt{r1{-}rj_{nd}} =(n_e(j),j=1,j_{nd})$; $\mathtt{rj_{nd+1}{-}rj_{nd+nt}} =(T_e(j),j=1,j_{nt})$;  $\mathtt{rj_{nd+nt+1}{-}rj_{ nd+nt+nt*nd}} =((C(j,j'),j'=1,j'_{nd}),j=1,j_{nt})$ (s$^{-1}$); $\mathtt{rj_{nd+nt+nt*nd+1}}=\lambda$~(\AA);  $\mathtt{i1} = nd$; $\mathtt{i2} = nt$; $\mathtt{i3} = i$ (lower level); $\mathtt{i4} = k$ (upper level); $\mathtt{i5}=Z$; $\mathtt{i6}=ion_N$.
\item[81.] Collision strengths for Fe~{\sc xix} \citep{bha89}: $\mathtt{r1} = \Upsilon(k,i)$; $\mathtt{i1} = i$ (lower level);  $\mathtt{i2} = k$ (upper level); $\mathtt{i3}=Z$; $\mathtt{i4}=ion_N$.
\item[82.] Decay rates for Fe UTA \citep{gu06}: $\mathtt{r1} =\lambda$~(\AA); $\mathtt{r2} =E(k)$~(eV); $\mathtt{r3} = gf(i,k)$; $\mathtt{r4} = A_r(k,i)$~(s$^{-1}$); $\mathtt{r5} = A_a(k,i)$~(s$^{-1}$); $\mathtt{i1} = i$ (lower level); $\mathtt{i2}=k$ (upper level); $\mathtt{i4}=ion_N$.
\item[83.]  Level data for Fe UTA \citep{gu06}: $\mathtt{r1} =E(i)$~(eV); $\mathtt{r2} = (2J+1)$; $\mathtt{r3} = 0.0$; $\mathtt{r4} =0.0$; $\mathtt{i1} = 1$; $\mathtt{i5}=i$ (level); $\mathtt{i6}=ion_N$; $\mathtt{s1}=$ level configuration assignment.
\item[85.] Photoionization cross sections for Fe ions obtained by summation of resonances near the K edge \citep{pal02}:  $\mathtt{r1} =Z_\mathrm{eff}$;  $\mathtt{r2} =E_{th}$~(Ryd);  $\mathtt{r3} =f$;  $\mathtt{r4} =\gamma$;  $\mathtt{r5} =$ scaling factor;  $\mathtt{i1} = n$; $\mathtt{i2}=L$; $\mathtt{i3}=2J$; $\mathtt{i4}=Z$; $\mathtt{i5}=k_{N-1}$; $\mathtt{i6}=ion_{N-1}$; $\mathtt{i7}=i_N$; $\mathtt{i8}=ion_N$.
\item[86.] Auger and radiative widths of $k_N$th K-vacancy level: $\mathtt{r1} =E(k_N)$~(eV, relative to $E(\infty)$); $\mathtt{r2} = A_a(k_N)$~(s$^{-1}$); $\mathtt{r3} = A_a(k_N,i_{N-1})$~(s$^{-1}$); $\mathtt{r4} = A_r(k_N)$~(s$^{-1}$); $\mathtt{i1} = i_{N-1}$; $\mathtt{i2}=k_N$; $\mathtt{i3}=Z$; $\mathtt{i4}=ion_{N-1}$; $\mathtt{i5}=ion_N$.
\item[88.]  Photoionization cross section damped excess of $i_N$th level of the $N$-electron ion leaving the ($N-1$)-electron ion in superlevel\_[K] $k_{N-1}$: $\mathtt{r1{-}rj_{max}} =(E(j),\sigma(E(j)),j=1,j_\mathrm{max})$~(Energy in Ryd relative to $E(\infty)$, cross section in Mb); $\mathtt{i1} = n$; $\mathtt{i2}=L$; $\mathtt{i3}=2J$; $\mathtt{i4}=Z$; $\mathtt{i5}=k_{N-1}$; $\mathtt{i6}=i_N$; $\mathtt{i7}=ion_N$.
\item[91.]  APED line ($k-i$) radiation rates \citep{fos12}: $\mathtt{r1} =\lambda$~(\AA); $\mathtt{r2} = 0.0$; $\mathtt{r3} = A(k,i)$~(s$^{-1}$); $\mathtt{i1} = i$ (lower level); $\mathtt{i2}=k$ (upper level); $\mathtt{i3}=Z$; $\mathtt{i4}=ion_N$.
\item[92.] APED collision strengths \citep{fos12}: $\mathtt{r1{-}rj_{max}} =(T_e(j),j=1,j_{max})$ (K); $\mathtt{rj_{max+1}{-}rj_{2*max}} =(\Upsilon(j),j=1,j_{max})$; $\mathtt{i1} = 1$; $\mathtt{i2}=i$  (lower level); $\mathtt{i3}=k$ (upper level); $\mathtt{i4}=Z$; $\mathtt{i5}=ion_N$.
\item[95.] Collisional ionization rates for $N$-electron ion \citep{bry06}:  $\mathtt{r1} =E(th)$ (eV);  $\mathtt{r2} =T_0$ (K); $\mathtt{r3{-}rj_{max+2}} =(\rho(j), j=1,j_{max})$ (effective collision strength); $\mathtt{i1} = i$ (level); $\mathtt{i5}=ion_N$.
\item[98.] Electron-impact effective collision strengths for the $k - i$ transition of the $N$-electron ion (CHIANTI fit \citep{bur92,der97}): $\mathtt{r1} =\Delta E$ (Ryd); $\mathtt{r2} =C$; $\mathtt{r3{-}rj_{max+2}} =(\Upsilon_{\rm red}(j),j=1,max)$ (reduced effective collision strength); $\mathtt{i1} = it$ (transition type); $\mathtt{i2} = i$; $\mathtt{i3} = k$; $\mathtt{i4}=ion_N$.
\item[99.]  Coefficients for recombination and photoionization cross sections of superlevels: $\mathtt{r1{-}rj_{nd}} =(n_e(j),j=1,j_{nd})$; $\mathtt{rj_{nd+1}{-}rj_{nd+nt}} =(T_e(j),j=1,j_{nt})$; $\mathtt{rj_{nd+nt+1}{-}rj_{ nd+nt+nt*nd}} =((\alpha(j,j'),j'=1,j'_{nd}),j=1,j_{nt})$; $\mathtt{rj_{nd+nt+nt*nd+1}-rj_{nd+nt+nt*nd+2*nx}} =(E(j),\sigma(j),j=1,j_{nx})$; $\mathtt{i1} = nd$; $\mathtt{i2} = nt$; $\mathtt{i3} = nx$; $\mathtt{i4} = n$; $\mathtt{i5}=L$; $\mathtt{i6}=2S+1$; $\mathtt{i7}=Z$; $\mathtt{i8}=k_{N-1}$; $\mathtt{i9}=ion_{N-1}$; $\mathtt{i10}=i_N$; $\mathtt{i11}=ion_N$.
\end{description}
\pagebreak


\section{Ion Models}\label{B}

\begin{table}[H]
\caption{Number of energy levels included in the {\sc xstar} ion models. \label{elev}}
\centering
\begin{tabular}{lrrrrrrrrrrrrrrr}
\toprule
\diagbox{N}{Z} & 1 & 2 & 3 & 4 & 5 & 6 & 7 & 8 & 9 & 10 & 11 & 12 & 13 & 14 & 15 \\
\midrule
 1 & 33 & 33 & 33 & 33 & 33 & 33 & 33 & 33 & 33 & 33 & 33 & 33 & 33 & 33 & 33 \\
 2 &    & 46 & 32 & 32 & 32 & 56 & 58 &241 & 52 & 54 & 52 & 58 & 52 & 58 & 52 \\
 3 &    &    &  3 &  3 &  3 & 26 & 43 & 43 & 22 & 43 & 22 & 43 & 22 & 43 & 22 \\
 4 &    &    &    &  3 &  3 & 22 & 53 & 53 & 43 & 79 & 43 & 79 & 43 & 79 & 43 \\
 5 &    &    &    &    &  3 & 22 & 58 &163 & 53 &163 & 53 &163 & 53 &163 & 53 \\
 6 &    &    &    &    &    &  8 & 61 & 79 & 53 & 79 & 53 & 79 & 53 & 79 & 53 \\
 7 &    &    &    &    &    &    & 82 & 34 & 34 & 34 & 34 & 34 & 63 & 34 & 34 \\
 8 &    &    &    &    &    &    &    & 19 & 3  & 19 & 19 & 19 & 57 & 19 & 19 \\
 9 &    &    &    &    &    &    &    &    & 3  &118 &  7 &  7 &  7 &  7 &  7 \\
10 &    &    &    &    &    &    &    &    &    &  4 &  3 & 46 & 46 & 46 & 50 \\
11 &    &    &    &    &    &    &    &    &    &    &  3 & 24 &  4 & 60 & 42 \\
12 &    &    &    &    &    &    &    &    &    &    &    &  3 &  4 & 57 & 31 \\
13 &    &    &    &    &    &    &    &    &    &    &    &    &  3 & 18 & 55 \\
14 &    &    &    &    &    &    &    &    &    &    &    &    &    &  3 &  3 \\
15 &    &    &    &    &    &    &    &    &    &    &    &    &    &    &  3 \\
\midrule
\diagbox{N}{Z} & 16 & 17 & 18 & 19 & 20 & 21 & 22 & 23 & 24 & 25 & 26 & 27 & 28 & 29 & 30 \\
\midrule
 1 & 33 &  33 &  33 &  33 &  33 &  33 &  33 &  33 &  33 &  33 &  33 &  33 &  33 &  33 &  33 \\
 2 & 58 &  52 &  38 &  52 &  46 &  52 &  52 &  52 &  52 &  52 &  50 &  52 &  43 &  52 &  52 \\
 3 & 43 &  22 &  43 &  22 &  43 &  22 &  22 &  22 &  22 &  22 &  52 &  22 &  53 &  22 &  22 \\
 4 & 79 &  43 &  79 &  43 &  79 &  43 &  43 &  43 &  43 &  43 & 199 &  43 &  79 &  43 &  43 \\
 5 &163 &  53 & 163 &  53 & 163 &  53 &  53 &  53 &  53 &  53 & 551 &  53 & 163 &  53 &  53 \\
 6 & 79 &  53 &  53 &  53 &  53 &  53 &  53 &  53 &  53 &  53 & 623 &  53 &  53 &  53 &  53 \\
 7 & 34 &  34 &  34 &  34 &  34 &  34 &  34 &  34 &  34 &  34 & 744 &  34 &  34 &  34 &  34 \\
 8 & 19 &  19 &  19 &  19 &  29 &  19 &  19 &  19 &  19 &  19 & 639 &  19 &  67 &  19 &  19 \\
 9 &  7 &   7 &   7 &   7 &   7 &   7 &   7 &   7 &   7 &   7 & 340 &   7 & 117 &   7 &   7 \\
10 & 46 &  50 &  72 &  50 &  78 &  50 &  50 &  50 &  50 &  50 & 276 &  50 &  98 &  50 &  50 \\
11 & 60 &  42 & 208 &  42 & 208 &  42 &  42 &  42 &  42 &  42 &  85 &  42 & 216 &  42 &  42 \\
12 & 53 &  31 & 153 &  31 & 149 &  31 &  31 &  31 &  31 &  31 & 146 &  31 & 171 &  31 &  31 \\
13 & 76 &  55 & 208 &  55 & 208 &  55 &  55 &  55 &  55 &  55 &  55 &  55 & 235 &  55 &  55 \\
14 & 82 &  71 & 249 &  71 & 249 &  71 &  71 &  71 &  71 &  71 &  68 &  71 & 269 &  71 &  71 \\
15 & 31 &  55 & 211 &  55 & 211 &  55 &  55 &  55 &  55 &  55 &  78 &  55 & 219 &  55 &  55 \\
16 &  3 &   3 & 143 &  31 & 143 &  31 &  31 &  31 &  31 &  31 &  70 &  31 & 145 &  31 &  31 \\
17 &    &   3 &  62 &  10 &  62 &  10 &  10 &  10 &  10 &  10 &  62 &  10 &  63 &  10 &  10 \\
18 &    &     &   3 &   3 &   3 &  54 &  54 &  54 &  54 &  54 &  31 &  54 &  40 &  54 &  54 \\
19 &    &     &     &   3 &   5 & 221 & 221 & 221 & 221 & 221 & 170 & 221 & 143 & 221 & 221 \\
20 &    &     &     &     &   3 &  32 & 258 & 258 & 258 & 258 &  54 & 258 & 258 & 258 & 258 \\
21 &    &     &     &     &     &   3 &  84 & 441 & 441 & 441 &  55 & 441 & 441 & 441 & 441 \\
22 &    &     &     &     &     &     &   3 &   3 & 537 & 537 &  33 & 537 & 537 & 537 & 537 \\
23 &    &     &     &     &     &     &     &   3 & 163 & 463 &  70 & 463 & 463 & 463 & 463 \\
24 &    &     &     &     &     &     &     &     &   3 & 160 &  73 & 295 & 295 & 295 & 295 \\
25 &    &     &     &     &     &     &     &     &     &   3 & 197 & 128 & 128 & 128 & 128 \\
26 &    &     &     &     &     &     &     &     &     &     &   3 &   3 &  40 &  40 &  40 \\
27 &    &     &     &     &     &     &     &     &     &     &     &   3 &   3 &   3 &   3 \\
28 &    &     &     &     &     &     &     &     &     &     &     &     &   3 &   3 &   3 \\
29 &    &     &     &     &     &     &     &     &     &     &     &     &     &   3 &   3 \\
30 &    &     &     &     &     &     &     &     &     &     &     &     &     &     &   3 \\
\bottomrule
\end{tabular}
\end{table}

\section{Atomic Data Provenance}\label{C}

\begin{table}[H]
\small
\caption{Data provenance including the {\sc chianti} \citep{der97}, PP95 \citep{pra95}, and AMDPP \citep{bad03,bad06} compilations. rt = rate type. dt = data type. \label{prov}}
\centering
\begin{tabular}{llllll}
\toprule
\textbf{ID} & \textbf{Code} & \textbf{rt} & \textbf{dt} & $\mathbf{(Z;N)}$ & \textbf{References} \\
\midrule
  ac81&1000&3&56& $(8;1)$  & \citep{abu81}\\
  ac92&1001&3&56& $(2;1)$  & \citep{agg92}\\
  ak91&1002&3&56& $(6;1)$  & \citep{agg91a}\\
  ak91a&1003&3&56& $(10;1)$ & \citep{agg91b}\\
  ak92&1004&3&56& $(14;1)$ & \citep{agg92a}\\
  ak92a&1005&3&56& $(20;1)$ & \citep{agg92b}\\
  ak93&1006&3&56& $(26;1)$ & \citep{agg93}\\
  ba00&1007&4&50& $(28;2)$ & \citep{bau00}\\
  bm98&1008&3&56& $(26;15)$ & \citep{bin98}, \textsc{chianti} (\citep{flo77,tay87}) \\
  bn00  &1009&3&56& $(26;20)$ & \citep{ber00}, \textsc{chianti} (\citep{kee87}) \\
  bp98&1010&3&56& $(26;25)$ & \citep{bau98} \\
  ch00&1011&3&51& $(10;7)$, $(20;8)$, $(26;16)$, $(28;14)$ & \textsc{chianti} (\citep{gau77}) \\
  ch00&1012&4&50& $(6,7,8;4)$, $(12,14;5)$,$(20;8)$, & \textsc{chianti} (\citep{gau77}) \\
      &    &  &  & $(26;6,12,14{-}18)$, $(28;14)$ & \\
  ch01&1013&3&51& $(6,7;3)$ & \textsc{chianti} (\citep{gau77}) \\
  ch01&1014&4&50& $(6,7;3)$ & \textsc{chianti} (\citep{wie66, mar93}) \\
  ch02&1015&4&50& $(8,10,12,14,16,18,20,26,28;3)$ & \textsc{chianti} (\citep{zha90,mar93})  \\
  ch02&1016&3&51& $(8,10,12,14,16,18,20,28;3)$ & \textsc{chianti} (\citep{zha90}) \\
  ch04&1017&3&51& $(6;4)$ & \textsc{chianti} (\citep{ber85,ber89})  \\
  ch04&1018&4&50& $(10,12,16;4)$ & \textsc{chianti} (\citep{mue76, sam84,zha92}) \\
  ch05  &1019&3&51& $(7;4)$ & \textsc{chianti} (\citep{ram94}) \\
  ch05&1020&4&50& $(14,18,26;4)$ & \textsc{chianti} (\citep{gau77, sam84}) \\
  ch06&1021&3&51& $(8;4)$ & \textsc{chianti} (\citep{kat90, zha92}) \\
  ch06&1022&4&50& $(20,28;4)$ & \textsc{chianti} (\citep{mue76, zha92}) \\
  ch07&1023&3&51& $(10,12,14,16,18,20,26,28;4)$ & \textsc{chianti} (\citep{sam84,zha92}) \\
  ch07&1024&4&50& $(6;5)$ &  \textsc{chianti} (\citep{dan78, nus81,len85}), \textsc{nist} \\
  ch08&1025&3&51& $(6;5)$ & \textsc{chianti} (\citep{blu92}) \\
  ch08&1026&4&50& $(7;5)$ & \textsc{chianti} (\citep{nus79,sta93}) \\
  ch09&1027&4&50& $(8,10;5)$ & \textsc{chianti} (\citep{dan78})  \\
  ch09&1028&3&51& $(7;5)$ & \textsc{chianti} (\citep{sta94a,sta94b}) \\
  ch10&1029&4&50& $(16,18,20;5)$ & \textsc{chianti} (\citep{dan78,bha86}) \\
  ch10&1030&3&51& $(8,10,12,14,16,18,20,26;5)$ & \textsc{chianti} (\citep{zha94}) \\
  ch11&1031&3&51& $(28;5)$ & \textsc{chianti} (\citep{sam86})  \\
  ch11&1032&4&50& $(26;5)$ & \textsc{chianti} (\citep{sam90}) \\
  ch12&1033&4&50& $(28;5)$ & \textsc{chianti} (\citep{dan78,bha86,sam86}) \\
  ch12&1034&3&51& $(7;6)$ & \textsc{chianti} (\citep{sta94a,sta94b}) \\
  ch13&1035&4&50& $(7;6)$ & \textsc{chianti} (\citep{gau77,bel95}) \\
  ch13&1036&3&51& $(8;6)$ & \textsc{chianti} (\citep{agg83,agg85,bha93,len94}) \\
  ch14&1037&4&50& $(8;6)$ & \textsc{chianti} (\citep{bha93}) \\
  ch14&1038&3&51& $(10;6)$ & \textsc{chianti} (\citep{agg84,bha93b,len94}) \\
  ch15&1039&4&50& $(10,12,14;6)$ & \textsc{chianti} (\citep{bha93b,bha93c}) \\
  ch15&1040&3&51& $(12;6)$ & \textsc{chianti} (\citep{bha95}) \\
  ch16&1041&3&51& $(14;6)$ & \textsc{chianti} (\citep{bha93b,bha93c} \\
  ch16&1042&4&50& $(16;6)$ & \textsc{chianti} (\citep{bha87},) \\
  ch17&1043&4&50& $(18;6)$ & \textsc{chianti} (\citep{der79} \\
  ch17&1044&3&51& $(16;6)$ & \textsc{chianti} (\citep{mas78,bha87})  \\
  ch18&1045&3&51& $(18;6)$ & \textsc{chianti} (\citep{der79}) \\
  ch18&1046&4&50& $(20;6)$ & \textsc{chianti} (\citep{fro85}) \\

\bottomrule
\end{tabular}
\end{table}

\setcounter{table}{1}
\begin{table}[H]
\small
\caption{-- \textit{continued}}
\centering
\begin{tabular}{llllll}
\toprule
\textbf{ID} & \textbf{Code} & \textbf{rt} & \textbf{dt} & $\mathbf{(Z;N)}$ & \textbf{References} \\
\midrule
  ch19&1047&3&51& $(20;6)$ & \textsc{chianti} (\citep{agg91c,agg91d})  \\
  ch19&1048&4&50& $(12,14,16,18,20;7)$ & \textsc{chianti} (\citep{bha80}) \\
  ch20&1049&4&50& $(26;7)$ & \textsc{chianti} (\citep{bha89b}) \\
  ch20&1050&3&51& $(26;6)$ & \textsc{chianti} (\citep{mas79,agg91e})  \\
  ch21&1051&4&50& $(14,16,18;8)$ & \textsc{chianti} (\citep{bha79})  \\
  ch21&1052&3&51& $(8;7)$ & \textsc{chianti} (\citep{mcl94})  \\
  ch22&1053&3&51& $(12,14,16,18,20;7)$ & \textsc{chianti} (\citep{bha80})  \\
  ch22&1054&4&50& $(26;8)$ & \textsc{chianti} (\citep{lou85}) \\
  ch23&1055&3&51& $(26;7)$ & \textsc{chianti} (\citep{bha80,bha89b}) \\
  ch24&1056&4&50& $(12,14,16,20;9)$ & \textsc{chianti} (\citep{bla94}), \textsc{nist} \\
  ch24&1057&3&51& $(10,12;8)$ & \textsc{chianti} (\citep{but94}) \\
  ch25&1058&3&51& $(14,16,18;8)$ & \textsc{chianti} (\citep{bha79},) \\
  ch25&1059&4&50& $(26,28;9)$ & \textsc{chianti} (\citep{sam91, cor92, bla94})  \\
  ch26&1060&4&50& $(14;10)$ & \textsc{chianti} (\citep{bha85}) \\
  ch26&1061&3&51& $(26;8)$ & \textsc{chianti} (\citep{lou85})  \\
  ch27&1062&4&50& $(16;10)$ & \textsc{chianti} (\citep{hib93}) \\
  ch27&1063&3&51& $(10;9)$ & \textsc{chianti} (\citep{sar94}) \\
  ch28&1064&3&51& $(12;9)$ & \textsc{chianti} (\citep{moh88}) \\
  ch28&1065&4&50& $(18,20,28;10)$ & \textsc{chianti} (\citep{zha87b,hib93}) \\
  ch29&1066&4&50& $(26;10)$ & \textsc{chianti} (\citep{bha92}) \\
  ch29&1067&3&51& $(14;9)$ & \textsc{chianti} (\citep{moh90,sar94}) \\
  ch30&1068&3&51& $(16;9)$ & \textsc{chianti} (\citep{moh87,sar94})  \\
  ch30&1069&4&50& $(12,14,16,18,20,26,28;11)$ & \textsc{chianti} (\citep{wie66,sam90}) \\
  ch31&1070&4&50& $(14;12)$ & \textsc{chianti} (\citep{gau77, duf83})  \\
  ch31&1071&3&51& $(20,26;9)$ & \textsc{chianti} (\citep{mas75,sar94})  \\
  ch32&1072&3&51& $(14;10)$ & \textsc{chianti} (\citep{bha85}) \\
  ch32&1073&4&50& $(16,18,20,28;12)$ & \textsc{chianti} (\citep{chr86}), \textsc{nist} \\
  ch33&1074&4&50& $(14;13)$ & \textsc{chianti} (\citep{gau77, lan94}) \\
  ch33&1075&3&51& $(16;10)$ & \textsc{chianti} (\citep{moh90b}) \\
  ch34&1076&4&50& $(16;13)$ & \textsc{chianti} (\citep{bha80b,duf82}) \\
  ch34&1077&3&51& $(18,20,28;10)$ & \textsc{chianti} (\citep{zha87b}) \\
  ch35&1078&3&51& $(26;10)$ & \textsc{chianti} (\citep{bha92}) \\
  ch35&1079&4&50& $(26;13)$ & \textsc{chianti} (\citep{fis86}) \\
  ch36&1080&4&50& $(26;19)$ & \textsc{chianti} (\citep{czy66, faw89}), \textsc{nist} \\
  ch36&1081&3&51& $(12,14,16,18,20,28;11)$ & \textsc{chianti} (\citep{sam90})  \\
  ch37&1082&3&51& $(14;12)$ & \textsc{chianti} (\citep{duf89})  \\
  ch37&1083&4&50& $(26;20)$ & \textsc{chianti} (\citep{nus82}) \\
  ch38&1084&3&51& $(16,18,20,28;12)$ & \textsc{chianti} (\citep{chr86}) \\
  ch39&1085&3&51& $(14;13)$ & \textsc{chianti} (\citep{duf91}) \\
  ch40&1086&3&51& $(16;13)$ & \textsc{chianti} (\citep{bha80b,bha80c,duf82}) \\
  ch41&1087&3&51& $(26;18)$ & \textsc{chianti} (\citep{faw91,mas94}) \\
  ch42&1088&3&51& $(26;19)$ & \textsc{chianti} (\citep{czy66,pin89})  \\
  ch43&1089&3&51& $(26;14)$ & \textsc{chianti} (\citep{faw89b,mas94})  \\
  cj94&1090&3&56& $(1;1)$ & \citep{cal94b,agg91a} \\
  cp99&1091&4&50& $(26;21)$ & \citep{che99} \\
  cp99&1092&3&56& $(26;21)$ & \citep{che99} \\
  eg99&1093&3&56& $(26;11)$ & \citep{duf89, eis99a} \\
  ei99&1094&3&56& $(26;12)$ & \citep{gau77,eis99b} \\
  fe01&1095&4&50& $(26;25)$ & \citep{gar62,nus88,gir95} \\
  fe10&1096&3&56& $(26;17)$ & \citep{gau77,pel95} \\
  fe24&1097&3&56& $(26;3)$ & \citep{zha90,ber97} \\
  gr58&1098&4&50& $(26;23)$ & \citep{gar58} \\ 
  im&1099&3&63& $(7,12,16,18,28;1)$, $(7;2)$ & Impact parameter method \\
  kh84&1100&4&50& $(2;2)$ & \citep{kon84} \\
  km87&1101&3&56& $(10,14,1;2)$ & \citep{sam83,goe83,kee87b,zha87} \\
  kn89&1102&3&69& $(6;2)$ & \citep{kat89} \\  
\bottomrule
\end{tabular}
\end{table}

\setcounter{table}{1}

\begin{table}[H]
\small
\caption{-- \textit{continued}}
\centering
\begin{tabular}{llllll}
\toprule
\textbf{ID} & \textbf{Code} & \textbf{rt} & \textbf{dt} & $\mathbf{(Z;N)}$ & \textbf{Reference} \\
\midrule
  kn89b&1103&3&69& $(2,8,12,20;2)$ & \citep{goe83,sam83,kat89} \\
  kn89c&1104&3&69& $(26;2)$ & \citep{goe83,sam83,kat89} \\
  mg05&1105&4&50& $(12;8)$ & PP95 (\citep{kau86}), \citep{nus79b,nus81,men83} \\
  ne01 &1106&4&50& $(10;7)$ & \citep{gau77}, PP95 (\citep{zei82,kau86}) \\
  ne02 &1107&4&50& $(10;8)$ & \citep{gau77}, PP95(\citep{men83,kau86}) \\
  np96 &1108&4&50&$(26;24)$ & \citep{nah96b} \\
  ny01 &1109&4&50&$(7;7)$ & \citep{but84,bau99a,bau99b} \\
  ox01 &1110&4&50& $(8;6)$ & PP95 (\citep{moo85,kau86}), \textsc{chianti} (\citep{bha93}) \\
  ox02 &1111&4&50&$(8;7)$ & \citep{gau77}, PP95 (\citep{zei82, kau86}) \\
  ox03 &1112&4&50& $(8;8)$ & \citep{bau99a,bau99b}, PP95(\citep{men83,kau86} \\
  pb95 &1113&3&56&$(18,20;17)$ & \citep{pel95} \\
  pp01 &1114&4&50& $(6;6)$ & PP95 (\citep{nus81,nus79b,kau86}) \\
  pp01 &1115&3&56&$(6;6)$ & PP95(\citep{peq76,joh87,tho75}) \\
  pp02 &1116&3&56& $(7;7)$ & PP95(\citep{dop76,ber81}) \\
  pp03 &1117&3&56& $(8;8)$ & PP95(\citep{led76,ber81}) \\
  pp04 &1118&3&56& $(16;14)$ & PP95 (\citep{men83,joh87, joh87b}) \\
  pp05 &1119&4&50& $(18,20;16)$ & PP95(\citep{men83,kau86}) \\
  pp05 &1120&3&56& $(18;14)$, $(20;19)$ & PP95 (\citep{men83}) \\
  pp06 &1121&3&56& $(18;15)$ & PP95 (\citep{zei87}) \\
  pp06 &1122&4&50&$(10;9)$, $(20;17)$ & PP95 (\citep{kau86}) \\
  pp07 &1123&3&56& $(18,20;16)$ & PP95 (\citep{kru70}) \\
  pp07 &1124&4&50&$(16;14)$ & PP95 (\citep{men82a,kau86,hay86}) \\
  pp08 &1125&4&50& $(18;15)$ & PP95 (\citep{men82b,kau86}) \\
  pp09 &1126&4&50& $(18;17)$ & PP95 (\citep{men83}) \\
  pp10 &1127&4&50& $(20;19)$ & PP95 (\citep{fuh90}) \\
  pp11 &1128&4&50& $(18;14)$ & PP95(\citep{duf82,kau86}) \\
  rb96 &1129&3&56&$(16;15)$ &  \citep{ram96}, \textsc{chianti} (\citep{cai93}) \\
  sm00 &1130&3&51& $(26;13)$ & \citep{sto00} \\
  ss01 &1131&4&50& $(16;15)$ & \citep{gau77}  \\
  ss96 &1132&3&56& $(18,20;13)$ & \citep{sar96} \\
  tb93 &1133&4&50& $(6,7,8,10,12,14,16,18,20,26;2)$ & \citep{cun93} \\
  zh96 &1134&3&56&$(26;24)$ & \citep{zha96} \\
  zp97 &1135&3&56&$(26;23)$ & \citep{zha97} \\
  zs87 &1136&3&56&$(18,28;2)$ & \citep{zha87} \\
  bm01 &1137&3&56& $(26;3)$ & \citep{bau03} \\
  bm01 &1138&4&50& $(26;3)$ & \citep{bau03} \\
  bm02 &1139&3&56& $(26;4{-}10)$ & \citep{bau04} \\
  pm01 &1140&4&50& $(26;4{-}9)$ & \citep{pal03b} \\
  pm02 &1141&4&50& $(26;18{-}25)$ & \citep{pal03c} \\
  mk01 &1142&4&50& $(26;10{-}17)$ & \citep{men04} \\
  gm01 &1143&4&50& $(8;2{-}8)$ & \citep{gar05} \\
  gk01 &1144&4&50& $(7;1{-}7)$ & \citep{gar09} \\
  pm03 &1145&4&50& $(16;2{-}16)$ & \citep{pal06} \\
  pq01 &1146&4&50& $(10,12,14,16,18,20;2{-}Z)$  & \citep{pal08a} \\
  pq02 &1147&4&50& $(28;2{-}27)$  & \citep{pal08b} \\
  pq03 &1148&4&50& $(14;2{-}13)$  & \citep{pal11} \\
  pq04 &1149&4&50& $(9,11,15,17,19;2{-}(Z-1))$  & \citep{pal12} \\
       &    & &  & $(21{-}25,27,29,30;2{-}(Z-1))$    & \\
  mb01 &1150&3&56& $(9,11,15,17,19;3{-}(Z-2))$ & Unpublished \\
  fj01 &1151&3&56& $(26;2,3)$                 & \citep{fos12} \\
  fj01 &1152&4&50& $(26;2,3)$                 & \citep{fos12} \\
  gu01 &1153&4&50& $(26;11{-}25)$                 & \citep{gu06} \\
  gu01 &1154&13&6& $(26;11{-}25)$                 & \citep{gu06} \\       
\bottomrule
\end{tabular}
\end{table}

\pagebreak

\setcounter{table}{1}

\begin{table}[H]
\small
\caption{-- \textit{continued}}
\centering
\begin{tabular}{llllll}
\toprule
\textbf{ID} & \textbf{Code} & \textbf{rt} & \textbf{dt} & $\mathbf{(Z;N)}$ & \textbf{Reference} \\
\midrule

  kf01 &1155&5&25& $(3{-}5,21,29,30;2)$,   & \citep{kin96} \\
       &    & &  & $(6{-}8,16;(Z-3){-}Z)$,         & \\
       &    & &  & $(10,12,14,16,18;(Z-2){-}Z)$,   & \\
       &    & &  & $(26,28;(Z-2){-}Z)$,            & \\
       &    & &  & $(20;(Z-1){-}Z)$                & \\
  bb01 &1156& 5, 15 &95& $(1{-}30;1{-}Z)$               & \citep{bry06} \\
  tb01 &1157&7&53& $(1{-}30;1)$, $(21{-}25,27{-}30;2)$,  & Hydrogenic \\
       &    & &  & $(28;3{-}10,21{-}26,28)$   & \\
 tb02 &1158&7&53& $(2{-}20,26;2)$                   & \citep{fer87} \\
 tb03 &1159&7&53& $(6,7,8,10,12,14,16,18,20,26;3)$  & \citep{pea88} \\
 tb04 &1160&7&53& $(6,7,8,10,12,14,16,18,20,26;4)$  & \citep{tul90} \\
 tb05 &1161&7&53& $(6,7,8,10,12,14,16,18,20,26;5)$  & \citep{fer99} \\
 tb06 &1162&7&53& $(6,7,8,10,12,14,16,18,20,26;6)$  & \citep{luo89} \\
 tb07 &1163&7&53& $(7,8,10,12,14,16,18,20,26;7)$    & \citep{cun93} \\
 tb08 &1164&7&53& $(8,10,12,14,16,18,20,26;8)$      & \citep{cun93} \\
 tb09 &1165&7&53& $(10,12,14,16,18,20,26;9)$        & \citep{cun93} \\
 tb10 &1166&7&53& $(10,12,14,16,18,20,26;10)$       & \citep{hib94} \\
 tb11 &1167&7&53& $(14,18,20,26;11)$                & \citep{cun93} \\
 tb12 &1168&7&53& $(26;12)$                         & \citep{but93} \\
 tb13 &1169&7&53& $(26;13)$                         & \citep{men95} \\
 tb14 &1170&7&53& $(14,26;14)$                      & \citep{nah93,cun93} \\
 tb15 &1171&7&53& $(16;16)$, $(18;18)$, $(26;15{-}21)$ & \citep{cun93} \\
 tb16 &1172&7&53& $(26;22)$                         & \citep{bau96} \\
 tb17 &1173&7&53& $(26;23)$                         & \citep{bau97} \\
 tb18 &1174&7&53& $(26;24)$                         & \citep{nah96} \\
 tb19 &1175&7&53& $(26;25)$                         & \citep{nah94} \\
 tb20 &1176&7&53& $(26;26)$                         & \citep{bau97b} \\
 tb21 &1177&7&53& $(28;27)$                         & \citep{bau99a} \\
 vy01 &1178& 1, 7 &59& $(3,4,5,9,11,15,17,19,21{-}25;1{-}Z)$, & \citep{ver95} \\
      &    &      &  & $(27,29,30;1{-}Z)$, $(13;1,2,12,13)$,  & \\
      &    &      &  & $(14;14)$, $(16;16)$, $(18;18)$, & \\
      &    &      &  &  $(20;18,29,30)$, $(26;19{-}25)$, $(28;21{-}28)$ & \\
 kp01 &1179&7&53& $(26;3{-}26)$                 & \citep{pal02,bau03,bau04,kal04} \\
 gg01 &1180&7&53& $(6;3{-}6)$                     & \citep{has10,gat18a}\\
 gm01 &1181&7&53& $(8;3{-}8)$                     & \citep{gar05} \\
 gg02 &1182&7&49& $(8;6{-}8)$                     & \citep{gor13,gat13a,gat13b} \\
 gk01 &1183&7&53& $(7;3{-}7)$                     & \citep{gar09} \\
 wk01 &1184&7&49& $(10,12,14,16,18,20;3{-}10)$    & \citep{wit09} \\
 gg03 &1185&7&49& $(10;8{-}10)$                   & \citep{gor00,jue06,gat15} \\
 wk02 &1186&7&49& $(12,14,16,18,20;11{-}17)$      & \citep{wit11a} \\
 gg04 &1187&7&49& $(12;3{-}12)$                   & \citep{has14} \\
 wk03 &1188&7&49& $(13;3{-}11)$                   & \citep{wit13} \\
 wk04 &1189&7&49& $(28;3{-}20)$                   & \citep{wit11b} \\
 tr01 &1190&7&49& $(9;3{-}(Z-2))$, $(15,17,19;3{-}11)$ & \citep{pal16} \\
      &    & &  &  $(21{-}25,27,29,30;3{-}11)$  & \\
 tr02 &1191&7&49& $(15,17,19;12{-}(Z-2))$,      & \citep{men17} \\
      &    & &  & $(21{-}25,27,29,30;12{-}18)$  & \\
 tr03 &1192&7&49& $(21{-}25,27;19{-}(Z-2))$,        & \citep{men18} \\
      &    & &  & $(28;21{-}26)$, $(29,30;19{-}26)$  & \\
 ap01 &1193&8&1& $(3{-}30;2)$, $(15{-}30;14{-}Z)$ & \citep{ald73,ald76} \\
 ap01 &1194&8&7& $(3{-}5;2{-}Z)$, $(7,21,29,30;2)$,  & \citep{ald73,ald76} \\
      &    & &  & $(13{-}25,27{-}30;13{-}Z)$        & \\
 ad01 &1195&8&38& $(3{-}12;3{-}Z)$, $(13{-}30;3{-}13)$ & AMDPP (\citep{bad06}) \\
 ad01 &1196&8&39& $(3{-}13;3{-}Z)$, $(14{-}30;3{-}12)$,  & AMDPP (\citep{bad03})\\
      &    & &  & $(26,28;13{-}20)$, $(26;21{-}26)$ & \\
 ns01 &1197&8&22& $(13;13)$, $(14;13,14)$        & \citep{nus86} \\
 gt01 &1198&8&30& $(1{-}30;1)$                   & \citep{gou70} \\
\bottomrule
\end{tabular}
\end{table}
\pagebreak
\reftitle{References}





\end{document}